\documentclass[aps,prl,twocolumn,amsfonts,nobibnotes,superscriptaddress,showpacs]{revtex4-1}

\usepackage{dcolumn}% Align table columns on decimal point
\usepackage{amsmath}
\usepackage{amssymb}
\usepackage{graphicx}
\usepackage{bm}
\usepackage{enumerate}
\usepackage{color}

\newcommand{\expe}[1]{\left\langle #1 \right\rangle}

\begin{document}

\title{The spontaneous symmetry breaking in Ta$_{2}$NiSe$_{5}$ is structural in nature}

\vspace{2cm}

\author{Edoardo Baldini}
\affiliation{Department of Physics, Massachusetts Institute of Technology, 02139 Cambridge, Massachusetts, USA}

\author{Alfred Zong}
\affiliation{Department of Physics, Massachusetts Institute of Technology, 02139 Cambridge, Massachusetts, USA}

\author{Dongsung Choi}
\affiliation{Department of Electrical Engineering \& Computer Science, Massachusetts Institute of Technology, 02139 Cambridge, Massachusetts, USA}

\author{Changmin Lee}
\affiliation{Department of Physics, Massachusetts Institute of Technology, 02139 Cambridge, Massachusetts, USA}

\author{Marios H. Michael}
\affiliation{Department of Physics, Harvard University, 02138 Cambridge, Massachusetts, USA}

\author{Lukas Windgaetter}
\affiliation{Max Planck Institute for the Structure and Dynamics of Matter, Hamburg, Germany}

\author{Igor I. Mazin}
\affiliation{Department of Physics and Astronomy and Center for Quantum Materials, George Mason University, 22030 Fairfax, Virginia, USA}

\author{Simone Latini}
\affiliation{Max Planck Institute for the Structure and Dynamics of Matter, Hamburg, Germany}

\author{Doron Azoury}
\affiliation{Department of Physics, Massachusetts Institute of Technology, 02139 Cambridge, Massachusetts, USA}

\author{Baiqing Lv}
\affiliation{Department of Physics, Massachusetts Institute of Technology, 02139 Cambridge, Massachusetts, USA}

\author{Anshul Kogar}
\affiliation{Department of Physics, Massachusetts Institute of Technology, 02139 Cambridge, Massachusetts, USA}

\author{Yao Wang}
\affiliation{Department of Physics, Harvard University, 02138 Cambridge, Massachusetts, USA}

\author{Yangfan Lu}
\affiliation{Department of Physics, University of Tokyo, Bunkyo-ku, Tokyo 113-0033, Japan}

\author{Tomohiro Takayama}
\affiliation{Department of Physics, University of Tokyo, Bunkyo-ku, Tokyo 113-0033, Japan}
\affiliation{Max Planck Institute for Solid State Research, 70569 Stuttgart, Germany}

\author{Hidenori Takagi}
\affiliation{Department of Physics, University of Tokyo, Bunkyo-ku, Tokyo 113-0033, Japan}
\affiliation{Max Planck Institute for Solid State Research, 70569 Stuttgart, Germany}

\author{Andrew J. Millis}
\affiliation{Department of Physics, Columbia University, New York, NY 10027, USA}
\affiliation{Center for Computational Quantum Physics, The Flatiron Institute, 162 Fifth Avenue, New York, NY 10010, USA}

\author{Angel Rubio}
\affiliation{Max Planck Institute for the Structure and Dynamics of Matter, Hamburg, Germany}
\affiliation{Center for Computational Quantum Physics, The Flatiron Institute, 162 Fifth Avenue, New York, NY 10010, USA}
\affiliation{Nano-Bio Spectroscopy Group, Departamento de F\'isica de Materiales, Universidad del Pa\'is Vasco, 20018 San Sebast\'ian, Spain}

\author{Eugene Demler}
\affiliation{Department of Physics, Harvard University, 02138 Cambridge, Massachusetts, USA}

\author{Nuh Gedik}
\affiliation{Department of Physics, Massachusetts Institute of Technology, 02139 Cambridge, Massachusetts, USA}

\date{\today}

%\keywords{}

\maketitle

\textbf{The excitonic insulator is an electronically-driven phase of matter that emerges upon the spontaneous formation and Bose condensation of excitons. Detecting this exotic order in candidate materials is a subject of paramount importance, as the size of the excitonic gap in the band structure establishes the potential of this collective state for superfluid energy transport. However, the identification of this phase in real solids is hindered by the coexistence of a structural order parameter with the same symmetry as the excitonic order. Only a few materials are currently believed to host a dominant excitonic phase, Ta$_2$NiSe$_5$ being the most promising. Here, we test this scenario by using an ultrashort laser pulse to quench the broken-symmetry phase of this transition metal chalcogenide. Tracking the dynamics of the material’s electronic and crystal structure after light excitation reveals surprising spectroscopic fingerprints that are only compatible with a primary order parameter of phononic nature. We rationalize our findings through state-of-the-art calculations, confirming that the structural order accounts for most of the electronic gap opening. Not only do our results uncover the long-sought mechanism driving the phase transition of Ta$_2$NiSe$_5$, but they also conclusively rule out any substantial excitonic character in this instability.}

The excitonic insulator (EI) is an elusive state of matter proposed theoretically in 1965
\cite{keldysh1965possible, des1965exciton, halperin1968possible} and expected to exhibit many unusual properties, such as superfluid energy
transport \cite{snoke2002spontaneous}, electronic ferroelectricity \cite{batista2004intermediate}, and superradiant emission \cite{mazza2019superradiant}. In several ways the EI is analogous to a conventional superconductor, but pairing in an electron-hole rather than electron-electron channel. Similar to a superconductor, the EI is a many-body effect beyond the scope of non-interacting electron theory. Unlike conventional superconductivity, the EI develops entirely within the electronic subsystem,
driven by electron-electron interactions based on Coulomb repulsion rather than phonon exchange. However, the EI instability has the same symmetry as a structural phase transition, so the EI and structural order parameters are in general linearly coupled and  occur together \cite{jerome1967excitonic}. Because there is no symmetry distinction, the question whether the transition is excitonic or structural in nature is necessarily quantitative rather than qualitative, involving comparison of energy scales. A phase can be classified as predominantly excitonic on the basis of two theoretically-defined criteria: (i) the instability occurs in the electronic subsystem alone, at fixed ionic positions \cite{keldysh1965possible, des1965exciton}, and (ii) the symmetry breaking leads to the emergence of a pseudo-Goldstone collective mode (phason) with a much smaller energy than the Higgs-like mode (see Supplementary Note 1) \cite{mazza2019nature}.

\begin{figure*}[t]
	\begin{center}
		\includegraphics[width=1.8\columnwidth]{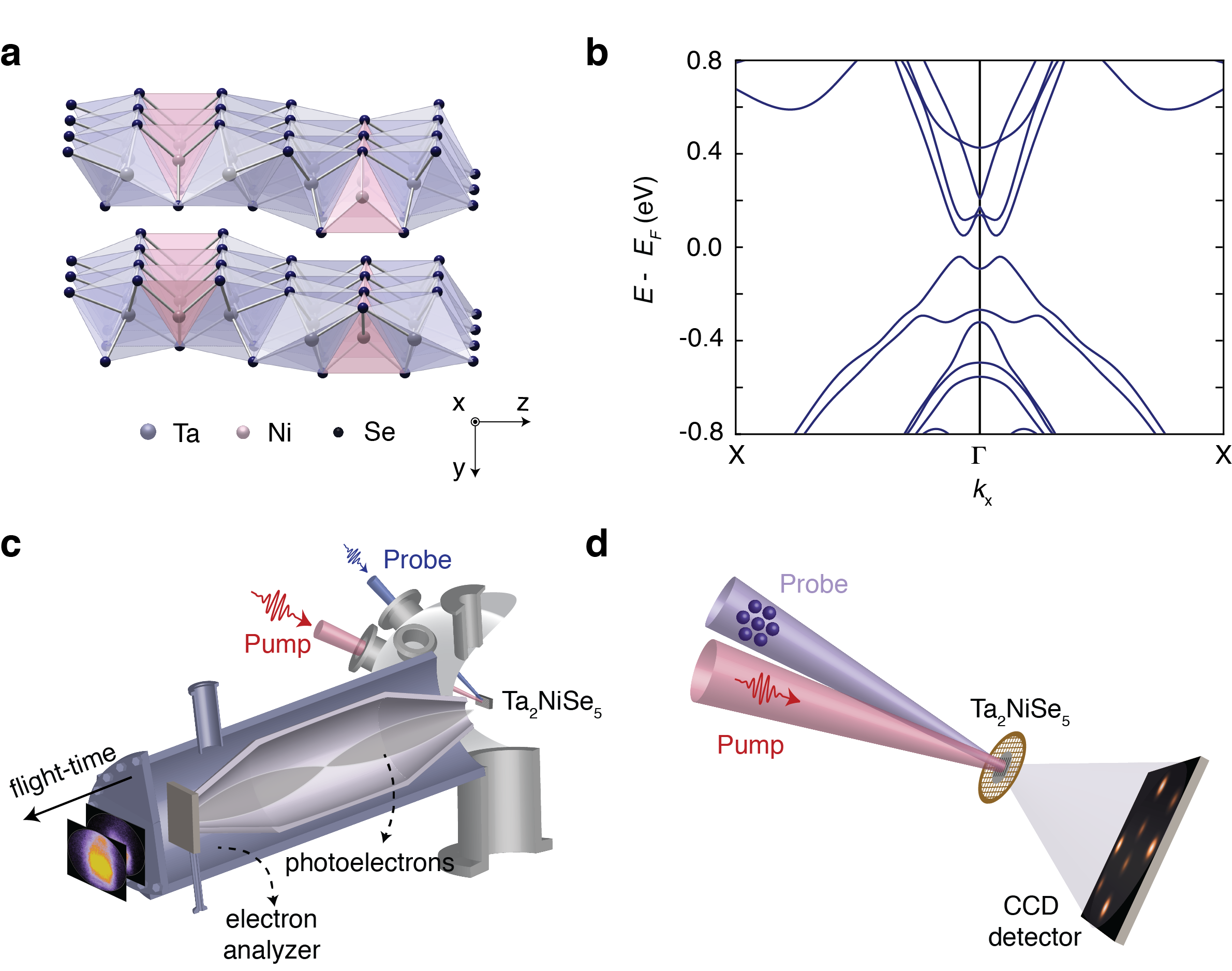}
		\caption{\textbf{Ta$_2$NiSe$_5$ and experimental methods.} \textbf{a}, High-temperature orthorhombic crystal structure of Ta$_2$NiSe$_5$ showing the layered nature of the system along the $y$ axis and the quasi-one dimensional Ta- and Ni chains running along the $x$ axis. The TaSe$_6$ octahedra and NiSe$_4$ tetrahedra are represented in light-blue and pink, respectively. \textbf{b}, Calculated electronic structure of Ta$_2$NiSe$_5$ in the low-temperature monoclinic unit cell along the X-$\Gamma$-X momentum direction (parallel to the Ta and Ni chains). The electronic structure is computed using $GW$ calculations. A bandgap opens in the single-particle band structure and its lower value is close to the $\Gamma$ point of the Brillouin zone (see Supplementary Note 7 for the details). The VB (CB) dispersions acquire an M-like (W-like) shape around $\Gamma$, consistent with the dispersion found in experiments. \textbf{c}, Schematic of the trARPES experiment. An ultrashort near-infrared pump pulse illuminates the sample and a delayed ultraviolet probe pulse photo-ejects electrons at different energies and momenta. The photoelectrons are finally detected in a time-of-flight analyzer. \textbf{d}, Schematic of the UED experiment performed in a transmission geometry. An ultrashort near-infrared pump pulse excites the sample and a delayed electron pulse is diffracted by the specimen and captured by a CCD detector. The specimen is in the form of an ultrathin flake deposited on a standard TEM Cu mesh.}
	\end{center}
\end{figure*}

There are only a few EI candidates, of which Ta$_{2}$NiSe$_{5}$ is one of the most extensively studied. Above a critical temperature $T_{C}$ = 328 K, this material crystallizes in a layered orthorhombic unit cell that consists of parallel Ta and Ni chains (Fig. 1a). At $T_{C}$, a second-order phase transition lowers the crystalline symmetry to monoclinic and the material simultaneously undergoes a semimetal-to-semiconductor transition \cite{lu2017zero,watson2019band}. Of note is the breaking of mirror
symmetry \cite{nakano2018antiferroelectric,mazza2019nature} and the development of spontaneous strain below $T_C$ \cite{di1986physical,nakano2018antiferroelectric}. In the semiconducting phase, a gap opens in the electronic structure \cite{lu2017zero,lee2019strong} and the valence band (VB) top acquires an M-like flat shape around the $\Gamma$ point of the Brillouin zone \cite{wakisaka2009excitonic,watson2019band}. Since this band flattening is expected from the Bogoliubov transformation for an electron-hole pair, it has been quoted as evidence of an electronic origin for the phase transition of Ta$_{2}$NiSe$_{5}$. In this scenario, the changes in the lattice degrees of freedom accompanying the electronic structure reconstruction are interpreted in terms of linear coupling of the lattice to the putative EI order parameter \cite{werdehausen2016coherent,mazza2019nature}. Nevertheless, it is crucial to remark that the opening of the hybridization gap and the M-shaped dispersion could also follow from a fundamentally distinct effect, the lowering of the crystal symmetry alone \cite{watson2019band,subedi}. Under this circumstance, the primary order parameter would be structural in nature, with relevant consequences on the fate of the EI. Such an underlying complexity in Ta$_2$NiSe$_5$ so far has posed significant challenges to disentangling different contributions to the gap formation in experiments performed under equilibrium conditions. This calls for the development of advanced nonequilibrium schemes to separate the time dependence of the electronic and structural components of the instability \cite{hellmann2012time,porer2014non,zhang2014ultrafast,zhang2020coherent}, assessing their relative importance with the support of state-of-the-art computational methods.

\begin{figure*}[t]
	\begin{center}
		\includegraphics[width=2\columnwidth]{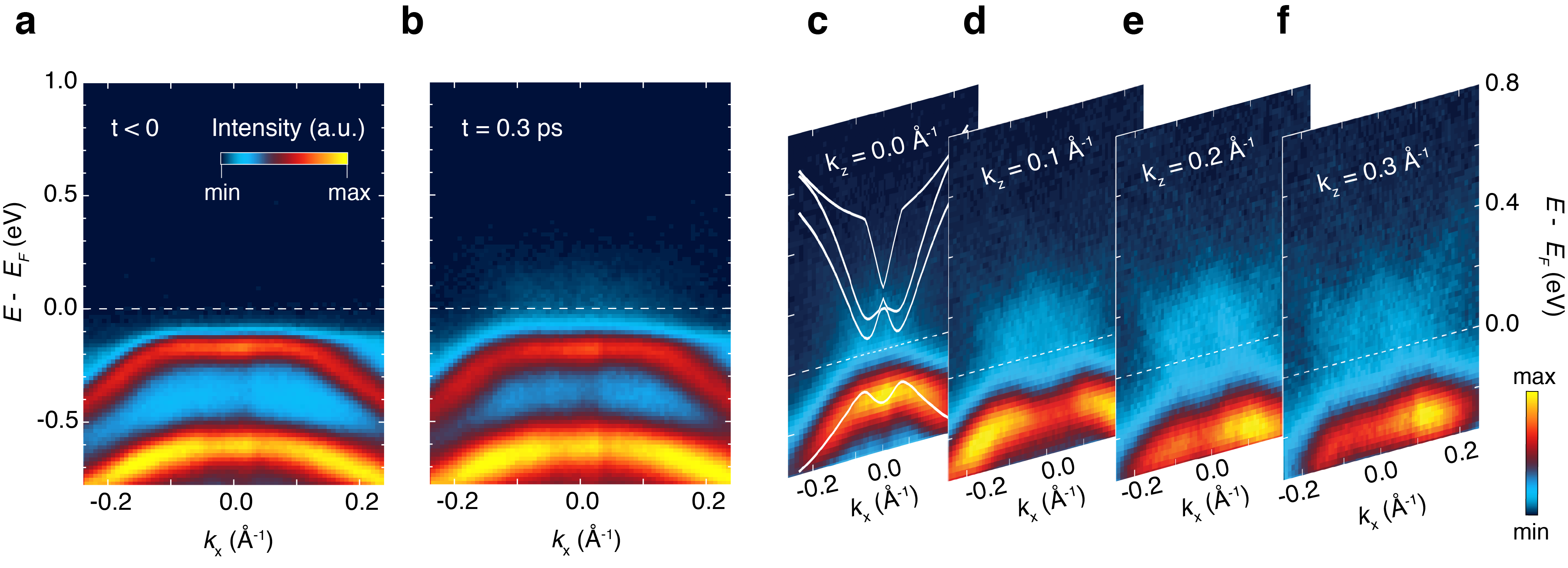}
		\caption{\textbf{Light-induced modification of the electronic structure.} \textbf{a,b} Snapshots of the trARPES spectra along the $k_x$ momentum direction and for $k_z$ = 0 \AA$^{-1}$. The data have been measured at 11 K with a probe photon energy of 10.75 eV and an absorbed pump fluence of 0.4 mJ/cm$^2$. (a) Snapshot before photoexcitation ($t$ $<$ 0). At the $\Gamma$ point of the Brillouin zone ($k_x$ = 0 \AA$^{-1}$), the flat anti-bonding VB is located around -0.16 eV, whereas the bonding VB appears around -0.65 eV. (b)~Snapshot measured at the maximum of the pump-probe response ($t$~=~0.3~ps). Upon photoexcitation, the VB is depleted in intensity and broadens significantly. Spectral weight is transfered above $E_F$ and accumulates close to $\Gamma$. \textbf{c-f}, Evolution of the photoexcited state (at $t$ = 0.3 ps) along $k_x$ at representative $k_z$ momenta, as indicated in the labels. Note that the color scale is different from that of panels (a,b). The spectral weight above $E_F$ assumes a W-like shape consistent with the dispersion of the CB. The VB and CB never crosses each other and thus the gap size remains finite in the whole $k_x$-$k_z$ momentum space around $\Gamma$. The white lines denote the energy-momentum dispersion calculated at the $GW$ level (Fig. 1b). A rigid shift of -84 meV has been applied to the VB to account for the underestimated gap resulting from the $GW$ method. The calculated dispersions have an excellent match with the experimental findings.}
	\end{center}
\end{figure*}

\begin{figure*}[t]
	\begin{center}
		\includegraphics[width=2\columnwidth]{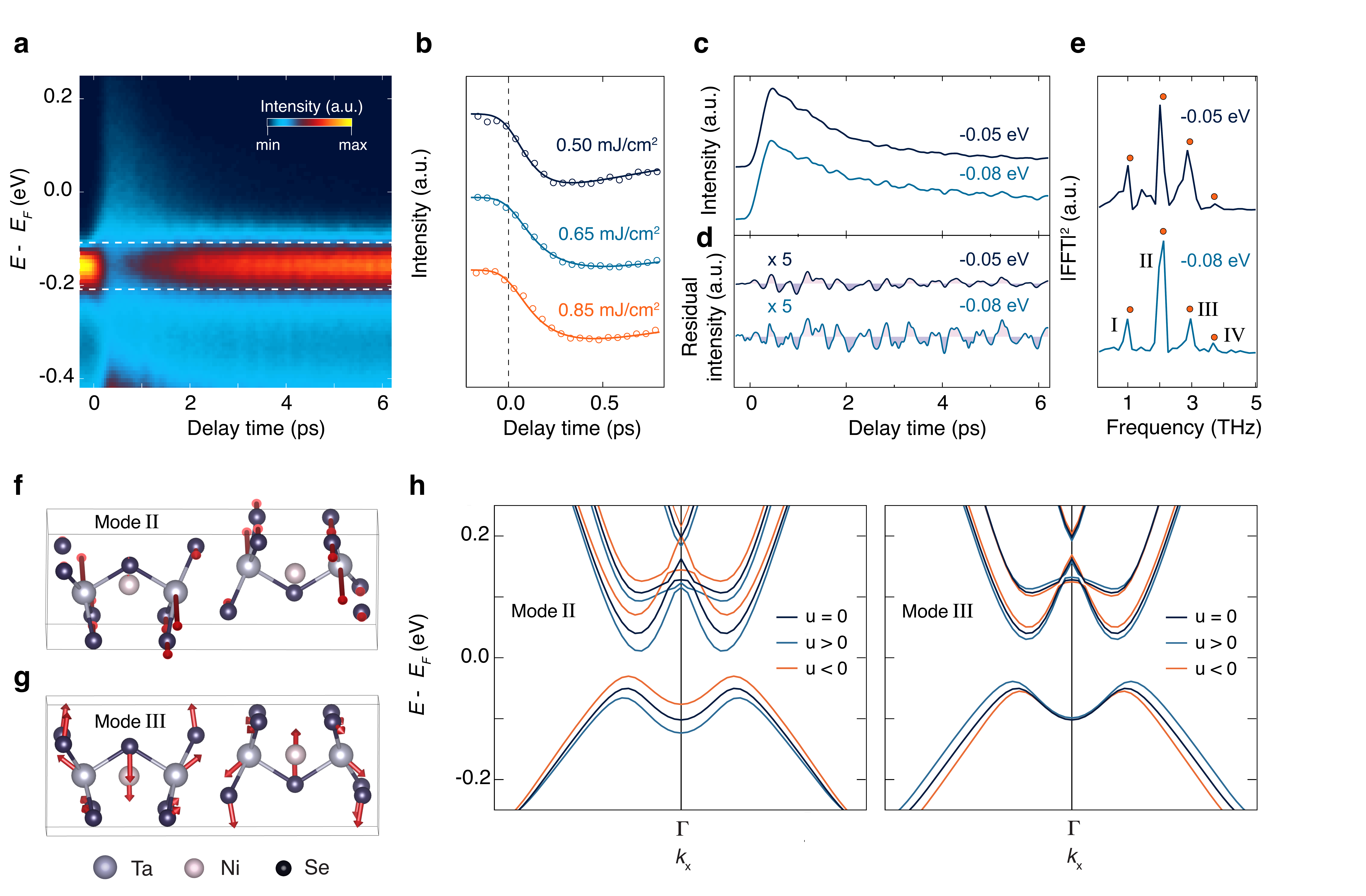}
		\caption{\textbf{Role of the collective modes in the gap response.} \textbf{a}, Map of the photoelectron intensity at the $\Gamma$ point as a function of energy and pump-probe delay. The data have been acquired at 14 K with a probe photon energy of 6.20 eV and an absorbed pump fluence of 0.85 mJ/cm$^2$. \textbf{b}, Excitation-density dependence rise of the photoelectron intensity response at $\Gamma$ (dotted lines). Fits to the experimental traces are overlapped in solid lines. The traces are selected at an energy of -0.158~eV with respect to $E_F$ and averaged over an energy window of $\pm$0.05 eV, as indicated by the dashed rectangle in panel (a). \textbf{c},~Time dependence of the momentum-integrated photoelectron intensity in selected energy intervals referenced to $E_F$. Intensities are normalized to the average intensity $I_0$ in the delay interval [-300,-50] fs; curves are offset for clarity and smoothed. The energy interval over which the intensity is integrated is $\pm$0.05 eV around the indicated energy. \textbf{d}, Oscillatory component singled out from the temporal traces of panel (c) by subtracting the nonoscillatory transient. For visualization purposes, the residuals have been multiplied by a factor of 5 and smoothed. \textbf{e}, Fourier transform analysis of the oscillatory signal in (d). Four frequency components (labelled as I-IV) appear in the spectrum and they are identified as Raman-active phonons of Ta$_2$NiSe$_5$. The corresponding frequencies detected in spontaneous Raman scattering \cite{werdehausen2016coherent} are indicated by orange dots. \textbf{f-g}, Calculated eigenvectors of the dynamical matrix of Ta$_2$NiSe$_5$ corresponding to modes II and III, respectively. Violet atoms refer to Ta, pink atoms to Ni, and blue atoms to Se. The phonon spectrum has been computed using DFT. To enhance the visibility of the atomic motion, the amplitude is scaled by a factor of 8. \textbf{h-i}, Calculated electronic structure of Ta$_2$NiSe$_5$ displaced along the eigenvectors of the modes showed in panels (f-g). The dark blue lines refer to the electronic structure of the initial (undisplaced, $u$ = 0) low-temperature unit cell, whereas the light blue (organge) lines indicate the band structure for positive (negative) displacements. The electronic structures are computed on the $GW$ level of theory.}
	\end{center}
\end{figure*}

Here, we present an experimental study of quench dynamics aimed at uncovering the nature of the phase transition in Ta$_{2}$NiSe$_{5}$. The ultrafast destabilization of the order parameter is imprinted on the material's nonequilibrium electronic
structure, which we track via time- and angle-resolved photoemission spectroscopy (trARPES, Fig.~1c). We observe that the electronic gap never collapses upon intense photoexcitation even though the depletion of the electronic states near  the VB maximum is strong; moreover, the dominant response proceeds over a phononic rather than electronic timescale. The central role of the crystal structure in the symmetry breaking is confirmed by direct visualization of the lattice dynamics, provided by ultrafast electron diffraction (UED, Fig.~1d). Advanced first-principles calculations performed in the realistic low-temperature unit cell clarify that the leading contribution to gap opening  in Ta$_{2}$NiSe$_{5}$ is of structural origin, as encountered in phonon-driven displacive transitions.

%showing that the changes in electronic structure track the changes in lattice structure rather than changes in the electronic distribution

As a first step, we map the light-induced modification of the electronic structure of Ta$_{2}$NiSe$_{5}$ via trARPES. We drive the material deep in the low-temperature phase ($T$~=~11 K) with an intense near-infrared laser pulse that rapidly changes the electronic distribution, in effect increasing the electronic temperature ($T_{e}$) to values well above $T_{C}$ while keeping the lattice cold. The theoretical estimates indicate that the carriers initially photoexcited by the pump rapidly relax, leading to a nonequilibrium distribution corresponding to a substantial depletion of the VB edge states (see Supplementary Note 2). The combination of a vacuum ultraviolet probe beam and a time-of-flight electron analyzer allows us to access a large portion of the Brillouin zone, elucidating how the electronic gap reacts to photoexcitation along both   $k_{x}$, the direction parallel to the chains in the orthorhombic cell and $k_{z}$, the one perpendicular to it. This feature is crucial because the material's electronic structure is not purely one-dimensional, with the interchain coupling establishing a well-defined dispersion along $k_{z}$ \cite{watson2019band}. If the instability in Ta$_{2}$NiSe$_{5}$ was purely excitonic in nature, our experimental protocol would lead to the observation of a complete gap closure on an electronic timescale.

\begin{figure*}[t]
	\begin{center}
		\includegraphics[width=1.8\columnwidth]{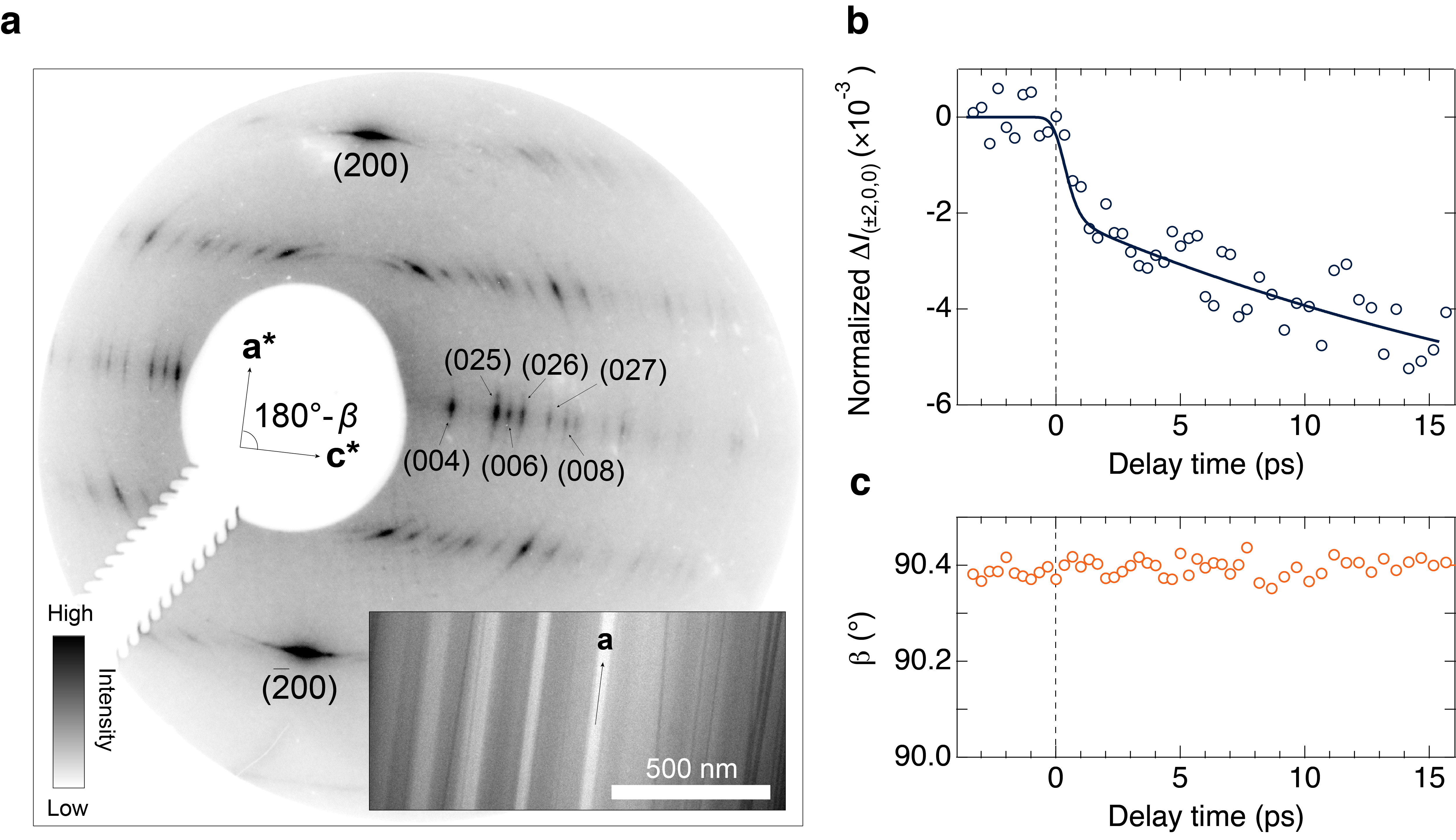}
		\caption{\textbf{Direct visualization of the structural dynamics.} \textbf{a},~Static electron diffraction pattern of Ta$_2$NiSe$_5$ taken at room temperature and with electrons at an energy of 26\,kV. Inset: electron micrograph of the UED sample taken at 120\,kV, showing nanoscale needle-like morphology. Though all needles are aligned along the $a$-axis, their $b$- and $c$-axis differ, resulting in peaks observed out of the [010] zone axis; some of these peaks are labeled. \textbf{b},~Photoinduced change of the integrated diffraction intensity, normalized to its value before excitation. The absorbed fluence is 0.1 mJ/cm$^2$. Intensity values are taken as the average between (200) and $(\bar{2}00)$ peaks. The fitted curve (solid line) shows a fast drop with $\tau_1=0.32\pm0.18$\,ps and a slow decay of $\tau_2=26\pm1$\,ps. \textbf{c},~Time evolution of the monoclinic distortion angle $\beta$, shown in panel (a), upon photoexcitation. While the sensitivity of the UED setup is not sufficient to resolve the pump-induced change, the angle never reaches the 90$^\circ$ value associated with the orthorhombic unit cell. Qualitatively similar data were measured at 77 K and 295 K.}
	\end{center}
\end{figure*}

Figures 2a,b show snapshots of trARPES spectra along $k_{x}$ for $k_{z}$ = 0 \AA$^{-1}$. At negative time delays (Fig.~2a), the flat anti-bonding VB is observed around an energy of -0.16 eV relative to the Fermi level ($E_F$) at the $\Gamma$ point, whereas the bonding VB appears at -0.65 eV \cite{wakisaka2009excitonic}. Photoexciting electron-hole pairs with $T_{e}$ of several hundred kelvins above $T_C$ induces a modification of the band structure that is the strongest around 0.3-0.4 ps (Fig.~2b). The flat VB is depleted in intensity and broadened substantially, but its peak energy remains nearly unchanged at all momenta (see Supplementary Figure 5). This is in stark contrast to the behavior observed in equilibrium upon increasing the lattice temperature, which involves a significant energy shift of the VB toward $E_F$ \cite{wakisaka2009excitonic,watson2019band}. Another important feature in the pump-probe spectrum is found above $E_{F}$, where spectral weight accumulates around the $\Gamma$ point. To investigate the nature of these states, we acquire data at 0.3 ps with improved sensitivity around $E_F$ and show them in Fig. 2c-f at representative $k_{z}$ values. At $k_{z}$ = 0 \AA$^{-1}$ and an energy of $\sim$50 meV (Fig.~2c), we observe an upward-dispersing band with a characteristic W shape, which we identify as the lowest conduction band (CB). While the exact estimate of the final gap size is hindered by the experimental resolution, the relevant aspect for our discussion is that the gap along $k_x$ remains open. Direct inspection of the snapshots simultaneously taken at finite $k_{z}$ (Fig. 2d-f) confirms that the VB and CB never cross each other throughout the two-dimensional momentum space around $\Gamma$. We remark that this behavior occurs in the presence of a clear separation between the electronic and lattice temperatures, differing from the response observed when the lattice is also transiently heated above $T_C$ \cite{okazaki2018photo, suzuki2020detecting}.

To elucidate whether the electronic or ionic degrees of freedom control the dynamics, we measure the trARPES signal with a high time resolution setup. Figure 3a displays the energy distribution of the photoemission intensity around $\Gamma$ (integrated over $\pm$0.05 \AA$^{-1}$ in the $k_x$-$k_z$ plane) as a function of pump-probe delay. Tracking the evolution of the VB intensity at different pump fluences (Fig.~3b) allows us to observe a response that is complete within $\sim$0.3-0.4 ps, a timescale longer than our instrument response function. This behavior is incompatible with the very short timescale (i.e. few fs, see Supplementary Note 5) that characterizes the plasma-induced screening of the Coulomb interaction. The latter effect plays an important role in the bandgap renormalization of conventional semiconductors and, in the case of a model two-band EI, it is expected to modify the excitonic gap amplitude on timescales of 10-100 fs \cite{golevz2016photoinduced,murakami2017photoinduced}. Direct inspection of our raw data and a combined analysis of the VB amplitude, broadening, and peak position indicates that plasma screening in Ta$_2$NiSe$_5$ provides only a small contribution to the gap renormalization compared to the phenomenon evolving on the slower timescale. The same analysis suggests that the dominant process involves the emission of optical phonons, a behavior usually observed in materials with a gap of structural origin \cite{schmitt2008transient,petersen2011clocking,hellmann2012time}. To assess the validity of this scenario, we search for an unambiguous signature of strongly-coupled phonon modes in Ta$_2$NiSe$_5$. Figure~3c displays the momentum-integrated photoemission response as a function of time at representative energies of -0.05 eV and -0.08 eV. Both curves refer to the upper edge of the VB and show the presence of oscillations in the photoemission intensity due to coherent phonons. Subtracting the multi-exponential background from the data of Fig. 3c allows us to isolate the signal of the collective response (Fig.~3d). The residuals reveal that the coherent phonon oscillations already emerge during the rise of the pump-probe signal, confirming that the maximum gap response is locked to a phononic timescale. Applying a Fourier filter to the signal of Fig.~3d yields the frequency spectrum of Fig.~3e. The peaks (labelled as I-IV) match the frequencies of four Raman-active phonons previously observed in other ultrafast studies \cite{werdehausen2016coherent, mor_inhibition, suzuki2020detecting, tang2020non, andrich2020imaging}. The sharp lineshape of mode II deserves special attention, as this is a characteristic fingerprint of the monoclinic phase of Ta$_2$NiSe$_5$ (see Supplementary~Note~6). It indicates that the crystal maintains the low-temperature structure even if the electronic distribution becomes strongly nonequilibrium, another feature incompatible with a symmetry breaking of purely excitonic origin.

To verify the inhibition of the monoclinic-to-orthorhombic transition, we directly visualize the evolution of the crystal structure through UED (details are given in the Methods). Figure\thinspace4a shows a static electron diffraction pattern of Ta$_{2}$NiSe$_{5}$ in the monoclinic phase, featuring two peaks at ($\pm2$,~0,~0). In Fig.~4b, we track the change in the diffraction intensity upon photoexcitation. We observe a signal that drops promptly and is followed by a slow evolution that marks the rise of lattice heating after $\sim$1~ps. Upon accounting for the instrument response function, we find that the initial sub-ps decrease is resolution-limited and indicates the immediate emission of multiple phonons (some of which revealed by trARPES) which redistribute intensity from Bragg reflections to elsewhere in the Brillouin zone. We then monitor how the $\beta$ angle associated with the monoclinic distortion reacts to the sudden increase of $T_e$. The results show that $\beta$ remains rather constant in time (Fig.~4c), never reaching the value of 90$^\circ$ expected for the orthorhombic cell. Although the sensitivity offered by our UED apparatus is not sufficient to resolve the small pump-induced change of $\beta$ (within a scale of 0.014$^\circ$), we can firmly establish that the structural transition never occurs at all time delays.

The presented data show that in Ta$_{2}$NiSe$_{5}$ the electronic distribution changes substantially and suddenly, the electronic gap remains open at all time delays, and the crystal retains its monoclinic structure. These results indicate that the material's instability is driven mostly by phonons and not by a pure EI order of electronic origin.

In the following, we rationalize these findings by performing state-of-the-art calculations based on DFT and its Hartree-Fock-like generalizations. As a first step, we establish the crystal structure that is favored at low temperature by relaxing the material's unit cell. We find that this structure is monoclinic, in agreement with previous results \cite{watson2019band,subedi,di1986physical}. We also study how the low-symmetry distortion reacts to an increase in $T_e$ by introducing a finite Fermi smearing in the Brillouin zone integrations. We observe that the electron heating does not remove the monoclinic distortion even at $T_{e}$$\gg$$T_{C}$, consistent with our experimental findings of Fig.~3e and Fig.~4c. The amplitude of the distortion is gradually reduced, albeit never to zero. Already at this stage, we could conclude that the orthorhombic-to-monoclinic transition is primarily driven by ion dynamics, and not by excitonic effects.

%While this is a minor problem in simple metals because the exchange interaction is exponentially screened, it becomes relevant for gapped solids. 

We then examine the evolution of the electronic structure as the lattice  is varied between the orthorhombic and monoclinic phases. To capture the electronic structure in the monoclinic phase, DFT alone is not sufficient, as the standard functionals routinely yield severely underestimated gaps and inaccurate band dispersions in semiconductors. Nevertheless, this fact is not related to any EI physics and stems from an incomplete description of the long-range exchange interaction in DFT \cite{maksimov1989excitation}. One can circumvent this problem by including the lowest-order correction in the screened electron-electron interaction. This is the essence of the so-called $GW$ method, which largely improves the description of semiconductors \cite{hedin1965new}. Importantly, this method does not account for any ladder-diagrammatic effects (such as the EI order) and
thus can serve as a crucial test for the EI hypothesis in Ta$_2$NiSe$_5$. We compute the material's electronic structure at the $GW$ level in the monoclinic unit cell and show the results in Fig. 1b (details are given in the Methods and in Supplementary Note~7). We observe that the monoclinic distortion alone profoundly affects the electronic structure, changing it from semimetallic to semiconducting.  Specifically, a hybridization gap opens around the $\Gamma$ point, its size being half of the experimental value \cite{lu2017zero}; the change in structure also leads to large systematic changes in band offsets, with CBs moving up in energy with respect to VBs. Further corrections in the description of the screening and the inclusion of the electron-phonon coupling would likely refine the gap size to larger values. More importantly, along $k_x$ the topmost VB acquires an M-like flat shape and the lowest CB develops a W-like structure, both consistent with the experimental dispersions (as shown by the white lines overlapped to the experimental data of Fig.~2c). These shapes are expected when two intersecting electron and hole bands hybridize, with the degree of flattening set by the strength of the hybridization potential. Starting from this electronic structure, we establish how it evolves upon increasing $T_e$. There are two mechanisms through which electron heating can modify the $GW$ gap: (i) Screening of the long-range exchange interaction, which proceeds on the electronic timescale set by the plasma frequency, and (ii) the dependence of the structural distortion (involving the monoclinic angle and the internal displacements) on $T_e$, an effect that evolves on phononic timescales. At our values of $T_e$, the former mechanism has a small impact on the electronic structure of Ta$_2$NiSe$_5$, consistent with our experimental findings (see Supplementary Note 7D). In contrast, a small decrease in the monoclinic distortion would cause a larger shrinkage of the $GW$ gap \cite{faleev2006finite}. In our trARPES data, the phononic timescale associated with the largest gap response reinforces the idea that the dynamics is governed by a small reduction of the structural distortion. During this time, the electron-phonon coupling plays a key role in equilibrating the electron and the ion subsystems. We can quantify this coupling for each of the coherent phonons emerging in trARPES by computing the $GW$ band structure while statically displacing the ions along the mode coordinates \cite{gerber2017femtosecond}. Figures~3f-i show representative results for modes II and III. In agreement with the experiment, our calculations show that the VB undergoes a substantial modulation around $\Gamma$, confirming a strong deformation potential coupling between the low-energy electronic states and the atomic displacements. Such deformation potential coupling involves, together with band offsets, strong changes in band hybridization.

In conclusion, our joint experimental-theoretical study excludes the scenario wherein the instability in Ta$_{2}$NiSe$_{5}$ has a dominant (or even substantial) EI nature. Rather, the electronic gap in the material can be best described as a one-electron hybridization gap driven by ion-ion interactions, with the possible addition of a small secondary EI order. In such conditions, the phason of the EI (if present) would be pinned at the high energy scale of the structural gap, hindering dissipationless transport and excitonic superfluidity even upon application of intense external stimuli \cite{mazza2019nature}. While our calculations (as well as those in Ref. \cite{subedi}) predict that a zone-center soft phonon leads to a quadrupolar order below $T_C$, only high-resolution structural probes will clarify the lattice dynamics responsible for the symmetry breaking. Indeed, anharmonic effects beyond the perturbative approach can also conspire to trigger the phase transition, making the experimental identification of the relevant phonon more challenging. Irrespective of the detailed mechanism at play, we believe that our study reconciles the controversial results obtained experimentally on Ta$_2$NiSe$_5$ since its discovery in 1985 \cite{sunshine1985structure}. We envision that the strategy presented here will serve as a general protocol to establish the role of the crystal structure in future candidate EIs.

\section*{Methods}

\subsection{Sample growth and preparation}

\label{Methods_SingleCrystals}

Single crystals of Ta$_{2}$NiSe$_{5}$ were synthesized by chemical vapour
transport. Elemental powders of Ta, Ni, and Se were mixed with
a stoichiometric ratio and sealed into an evacuated quartz tube ($\sim$1$\times$10$^{3}$ Pa) with a small amount of I$_{2}$ as transport agent. The
mixture was sintered under a temperature gradient of 950/850$^{\circ}$C.
After sintering for 1 week, needle-like single crystals grew at the cold
end of the tube.

\noindent For the trARPES experiment, the single crystals were directly glued on a copper holder using silver epoxy, in order to ensure a good thermal contact in the cryostat. For the UED experiments, an ultramicrotome fitted with a diamond
blade was used to cleave a single crystal of Ta$_{2}$NiSe$_{5}$ along the
$ac$ plane, producing thin flakes with an approximate dimension of 600\,$\mu$m~$\times$~50~nm~$\times$~200\,$\mu$m. Flakes were scooped from water onto standard transmission electron microscopy (TEM) copper grids (300\,lines/inch). The TEM grids were clamped to a copper
holder that ensures good thermal contact. Sample characterization was done by
a commercial TEM (Tecnai G2 Spirit TWIN, FEI) with a 120-kV electron beam energy.

\subsection{Time- and angle-resolved photoemission spectroscopy}

\label{Methods_Ultrafast}

The Ta$_{2}$NiSe$_{5}$ single crystals were cleaved at 10-14~K under
ultra-high-vacuum conditions ($<$1$\times$10$^{-10}$ torr). Systematic
trARPES data were reproduced on a total of ten samples using two different laser
schemes. The first scheme used a setup based on an amplified Yb:KGW laser system operating at 100 kHz (PHAROS SP-10-600-PP, Light Conversion). Details are reported in Ref. \cite{lee2019high}. In brief, the laser ouput (with pulses centered around 1.19~eV) was split into a pump and probe beams. The pump beam was directed into an optical parametric amplifier (ORPHEUS,
Light Conversion) to produce a near-infrared pulse at 1.55~eV. The probe pulse was frequency tripled to 3.58~eV and directed into a hollow fiber filled with Xe gas (XUUS, KMLabs). Here, pulses centered around 10.75 eV were obtained through nonlinear conversion of the 3.58~eV beam. The resulting
vacuum ultraviolet pulse was passed through a custom-built grating
monochromator (McPherson OP-XCT) to minimize pulse width broadening and enhance throughput efficiency. Finally, the probe was focused onto the sample with an in-plane polarization state perpendicular to the Ta and Ni chains. The temporal resolution of the setup was 230 fs, while the energy resolution was 43~meV. The second laser scheme consisted of an amplified Ti:Sapphire system (Wyvern, KMLabs), emitting ultrashort pulses around 1.55~eV and at a
repetition rate of 30 kHz. A portion of the output beam was used directly as the near-infrared pump pulse at 1.55~eV, whereas the ultraviolet probe was obtained by
frequency-quadrupling the laser fundamental photon energy to 6.20~eV. The
probe light polarization state was set to circular. The overall time resolution was
$\sim$160~fs (see Supplementary
Note 3), while the energy resolution was 31~meV. 

A time-of-flight analyzer (Scienta ARTOF 10k) was used to acquire the transient band structure of Ta$_{2}$NiSe$_{5}$ in the two-dimensional $k_x$-$k_z$ plane around the $\Gamma$ point of the Brillouin zone without rotating the sample or the detector. The pump beam was incident on the sample at an angle of $\sim$45$^{\circ}$ and its
polarization could be set precisely to either S or P with respect to the incident plane. The
sample was oriented such that S polarization had a pure in-plane electric
field component perpendicular to the chains (\textbf{$E_{in}$} $\parallel$ c),
while P polarization had an in-plane component parallel to the chains
(\textbf{$E_{in}$} $\parallel$ a) and an out-of-plane component perpendicular
to the sample surface (\textbf{$E_{out}$} $\parallel$ b). The position of $E_F$ was carefully calibrated for each sample by acquiring the steady-state ARPES spectrum of an auxiliary Bi$_2$Se$_3$ single crystal.

\subsection{Ultrafast electron diffraction}

The 1.19 eV output of an amplified Yb:KGW laser system
(PHAROS SP-10-600-PP, Light Conversion) operating at 100 kHz was split into
pump and probe branches. The pump beam was focused onto the sample, while the probe beam was frequency quadrupled to 4.78 eV and
focused onto a gold-coated sapphire in high vacuum ($<$4$\times$10$^{-9}$\,torr)
to generate photoelectrons. These electrons were accelerated to 26 kV in a dc
field and focused with a solenoid before diffracting from Ta$_{2}$NiSe$_{5}$
in a transmission geometry. Diffracted electrons were incident on an
aluminum-coated phosphor screen (P-46), whose luminescence was recorded by an
intensified charge-coupled device (iCCD PI-MAX II) operating in
shutter mode. The temporal resolution was 0.8\,ps (see Supplementary
Note 3). We calculated the $\beta$ angle by fitting the positions of the (200), (-200) and (004) Bragg peaks, which allows for the determination of the reciprocal unit vectors [100] and [001]. Small detector aberrations resulted in a $\beta$ angle of $\sim$90.4$^\circ$, slightly deviating from the 90.57$^\circ$ reported previously \cite{di1986physical}. The experiments were performed at 77 K and 300 K. 

\subsection{First-principles calculations}

All \textit{ab initio} calculations were performed using the Vienna Atomistic Simulation Package (VASP) implementing the projected augmented wave (PAW) method \cite{Vasp1}. The DFT structural relaxation using the vdw-opt88-PBE functional (know to correctly describe both van der Waals and regular interactions) resulted in the monoclinic structure characteristic of the low-temperature phase of Ta$_2$NiSe$_5$. The relaxation was performed on a 24$\times$4$\times$6 $k$-mesh with a 460 eV cutoff. We obtained the lattice parameters $a$ = 3.517~\AA, $b$ = 12.981~\AA, and $c$~=~15.777~\AA, and unit cell angles $\alpha$ = 90.005$^\circ$, $\beta$~=~90.644$^\circ$, and $\gamma$ = 89.948$^\circ$. Note that this structure has a small triclinic distortion, but it is very close to the monoclinic one, as analyzed in the Supplementary Note 7. This structure agrees, within our numerical accuracy, to the monoclinic $C2/c$ cell measured in experiments \cite{sunshine1985structure,nakano2018antiferroelectric} and in agreement with the findings of Ref. \cite{subedi}.

DFT band structure calculations were performed on a 16$\times$4$\times$4 $k$-mesh using the standard Perdew-Burke-Ernzerhof (PBE) exchange-correlation functional \cite{Perdew1996}. Afterwards, the $GW$ bands were calculated on top of DFT at the $G_0W_0$ level with a 12$\times$4$\times$2 $k$-mesh. We used a 100~eV cutoff for the calculation of the random-phase approximation (RPA) polarizability and we included 1086 conduction bands and 160 frequencies for the calculation of the screened interaction. For the analysis of the electronic temperature effect on the band structure, we performed $G_0W_0$ calculations with the parameters reported above on top of DFT calculation with the PBE functional. A Fermi smearing for different temperatures was set for the self-consistent cycle. This allowed us to have a depletion of the top valence bands and finite occupation of the bottom conduction bands, which in turns affected the $G_0W_0$ corrections as discussed in the text. The complete electronic structure analysis is given in Supplementary Note 7C,D.

The computation of the phonon dispersion and eigenmodes was performed on a 4$\times$4$\times$3 $k$-mesh with a 4$\times$2$\times$1 supercell. The analysis of the phonon modes was performed with the \textit{Phonopy} package \cite{phonopy}. To analyze the effect of representative phonon modes on the electronic bands, we performed $G_0W_0$ calculations as described above, but with the atomic positions displaced along the phonon eigenmodes by an amount equal to $\pm\sqrt{\langle {\bf r}^2\rangle_{\rm{T=0K}}}$, with $\langle {\bf r}^2\rangle_{\rm{T=0K}}$ being the amplitude of the zero-point oscillation of this mode. Similar calculations were repeated at different temperatures. More details are provided in Supplementary Note 7E,F.

\section{Data availability}

The data that support the findings of this study are available from the
corresponding author upon reasonable request.

\section{Supplementary Information}
Supplementary Information is available for this paper.

\section{Acknowledgements}
We thank F. Boschini, S. Kaiser, D. Chowdhury, A. Georges, G. Mazza, and A. Subedi for insightful discussions. We are grateful to F. Mahmood, E.J. Sie, T. Rohwer, B. Freelon for early instrumentation work of the trARPES and UED setups at MIT. The work at MIT was supported by DARPA DSO under DRINQS program grant number D18AC00014 (trARPES data taking and analysis), Army Research Office Grant No. W911NF-15-1-0128 (instrumentation for the trARPES setup), and the US Department of Energy BES DMSE (UED measurements). The work at Harvard was supported by Harvard-MIT CUA, AFOSR-MURI: Photonic Quantum Matter (award FA95501610323), and DARPA DRINQS program (award D18AC00014). E.B acknowledges additional support from the Swiss National Science Foundation under fellowships P2ELP2-172290 and P400P2-183842. Y.W. acknowledges the Postdoctoral Fellowship in Quantum Science of the Harvard-MPQ Center for Quantum Optics. The theory work was supported by the European Research Council (ERC-2015-AdG694097), the Cluster of Excellence ``Advanced Imaging of Matter" (AIM), Grupos Consolidados (IT1249-19) and SFB925. The Flatiron Institute is a division of the Simons Foundation. Support by the Max Planck Institute - New York City Center for Non-Equilibrium Quantum Phenomena is acknowledged. S. L. acknowledges support from the Alexander von Humboldt foundation. I.I.M. acknowledges support from the Office of Naval Research (ONR) through the grant \#N00014-20-1-2345. A.J.M. acknowledges e support from DOE BES Pro-QM EFRC (DE-SC0019443). Part of the calculations used resources of the National Energy Research Scientific Computing Center (NERSC), a US Department of Energy Office of Science User Facility operated under Contract No. DE-AC02-05CH11231. A.Z. and A.K. thank Y.~Zhang for assisting us at the MRSEC Shared Experimental Facilities at MIT, supported by the NSF under award number DMR-14-19807. A.Z. and A.K. also thank C.~Marks for the assistance in preparing UED samples at Center for Nanoscale Systems, a member of the National Nanotechnology Coordinated Infrastructure Network (NNCI), which is supported by the National Science Foundation under NSF award no. 1541959. CNS is part of Harvard University.

\section{Author Contributions}
E.B. conceived the project. E.B., D.C., D.A. and B.L performed the trARPES experiments with the 10.75~eV probe setup. E.B., C.L. and A.Z. performed the trARPES experiments with the 6.20 eV probe setup. A.Z. and A.K. performed the UED experiments. Y.L., T.T. and H.T. performed the crystal growth. E.B. and A.Z. analyzed the data. L.W., I.I.M, S.L. and A.R. performed the first-principles calculations. M.M., Y.W., I.I.M., A.J.M. and E.D. performed the analytical calculations. I.I.M., A.J.M., A.R., E.D. and E.B. interpreted the data. E.B., I.I.M., A.J.M., E.D. and N.G. wrote the manuscript. All the authors contributed to the final version of the paper. The entire project was supervised by N.G.

\section{Author Information}
The authors declare having no competing financial interests. Correspondence and requests for materials should be addressed to N.G. (email: gedik@mit.edu).
\newpage
\clearpage

\setcounter{section}{0}
\setcounter{figure}{0}
\renewcommand{\thesection}{S\arabic{section}}  
\renewcommand{\thetable}{S\arabic{table}}  
\renewcommand{\thefigure}{S\arabic{figure}} 
\renewcommand\Im{\operatorname{\mathfrak{Im}}}
%\titleformat{\section}[block]{\bfseries}{\thesection.}{1em}{} 

\section*{Supplementary Note 1: Fate of the excitonic insulator in the presence of phonons}

In this section, we discuss the fate of the phason of an EI state in the two limiting cases where the transition is driven either by an electronic instability or by a structural instability. A detailed analysis of an EI coupled to a phonon was recently published \cite{Murakami20}; here, as a reference for the rest of the paper, we provide a simplified analysis and a more general perspective. Starting from a simple two band model of an EI, the order parameter is given by the hybridization between the two bands $\Delta(x) = \expe{c_1^\dag(x) c_2(x)}$, where $c_1^\dag(x)$ creates an electron in band 1, $c_2(x)$ annihilates an electron in band 2, and $\expe{...}$ indicates the expectation value. The structural instability is described by an optical phonon coordinate $X_{phon}$. One can then derive a low-energy effective theory from this microscopic model. Using the symmetry of the phonon and the excitonic order parameter and expanding in the lowest order terms of the relevant fields (which should be justified not too far from the transition temperature $T_C$), we obtain the Lagrangian
\begin{widetext}
\begin{subequations}
	\begin{align}
	L =& \int d^2 x \left( - \frac{1}{2} \Delta^* \partial_t^2 \Delta + \frac{\xi}{2} \Delta^* \partial_x^2 \Delta - M_{phon}\frac{X_{phon} \partial_t^2 X_{phon}}{2 }- V(X_{phon} , \Delta)  \right) , \\
	V(X_{ph.} , \Delta) =& 
	\frac{r}{2} |\Delta|^2 + \frac{g}{4} |\Delta|^4 + \lambda X_{phon} \frac{(\Delta + \Delta^*)}{2} + \frac{\kappa}{2} X^2_{phon} + \frac{u}{4} X^4_{phon} + \frac{v}{2} X_{phon}^2 |\Delta|^2,
	\end{align}
	\label{eq:lang}
\end{subequations}
\end{widetext}
where $M_{phon}$ is the ionic mass associated with the phonon. The quartic terms $g$,~$u$,~$v$~$>$~0. The quadratic coefficients in the potential, $r$ and $\kappa$, respectively drive the excitonic and structural instability as they approach zero. Here, for simplicity, we neglect the U(1) symmetry breaking that, as noted by Mazza et. al. \cite{mazza2019nature}, is intrinsic to solid-state realizations of the EI phenomenon, on the grounds that this symmetry breaking is very small in practice. In the dynamics, we neglect dissipative terms arising from particle-hole excitations. As pointed out by Pekker and Varma \cite{Pekker15}, in this case the lowest-order time-derivative term for the order parameter is second order due to particle-hole symmetry. The theory is thus valid only for frequencies less than twice the quasiparticle gap.  Finally, $\lambda$ couples the two modes linearly.\\

\noindent \textbf{A. Phase diagram}\\

\begin{figure}
		\begin{center}
	\includegraphics[width=\columnwidth]{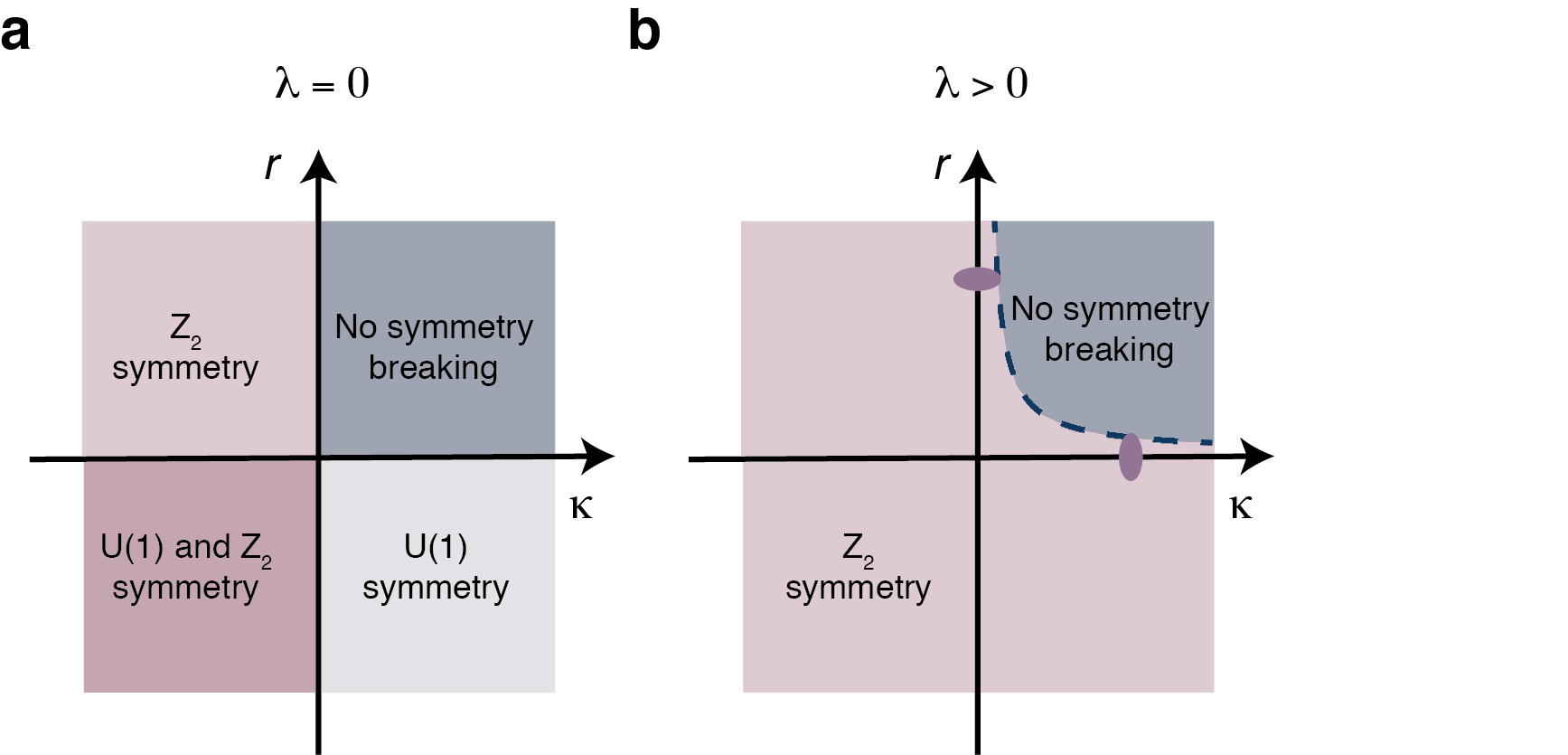}
		\end{center}
	\caption{\textbf{Phase diagram of the EI-phonon coupled system.} \textbf{a,} Phase diagram without coupling of electrons to phonons ($\lambda$ = 0). In this case, there is a U(1) symmetry in the electronic sector and a $Z_2$ symmetry in the structural sector which can be broken independently leading to four distinct phases. \textbf{b,} Phase diagram with finite electron-phonon coupling ($\lambda$ $>$ 0). U(1) symmetry is explicitly broken by linear coupling to the phonons, leading to the three symmetry-broken phases to merge into only one $Z_2$ phase. The purple dots signify the two limiting cases where the transition is driven by the excitonic order parameter or the crystal structure.}
	\label{fig:phase}
\end{figure}

For $\lambda = 0$, the above problem factorizes into a $U(1)~\times~Z_2$ symmetry sector. The phase diagram for this case is presented in Fig.~\ref{fig:phase}a and it comprises four distinct phases: i) $r$ $>$ 0 , $\kappa$ $>$ 0, with no symmetry breaking, ii) $r$ $<$ 0, $\kappa$ $>$ 0, an EI phase with U(1) symmetry breaking, iii) $r$ $>$ 0 , $\kappa$ $<$ 0, a phase with $Z_2$ symmetry breaking in the form of a structural phase transition, and iv) $r$ $<$ 0, $\kappa$ $<$ 0, with both U(1) and $Z_2$ symmetry breaking. However, for $\lambda >$ 0, the U(1) symmetry of the electron system is explicitly broken by linear coupling to the phonon, and all three symmetry-broken phases for $\lambda = 0$ merge into a single phase of spontaneous $Z_2$-symmetry breaking (shown in Fig.~\ref{fig:phase}b). First, we express the order parameter in terms of its real and imaginary parts: $\Delta = \Delta_1 + i \Delta_2$. Then, due to the presence of the phonon, we can set $\Delta_2 = 0$ and find the minimum of $\Delta_1$. The resulting equations are
\begin{subequations}
	\begin{align}
	r \Delta_1 + g \Delta_1^3 + \lambda X_{phon} + v X^2_{phon} \Delta_1 &= 0 , \\
	\kappa X_{phon} + u X_{phon}^3 + \lambda \Delta_1 + v X_{phon} \Delta_1^2 &=0.
	\end{align}
\end{subequations}
For small $\lambda$, we can perturbatively find the correction to the phase diagram near the two extreme scenarios $ |r|\ll$1, $\kappa$ $>$ 0 (corresponding to an electronically-driven transition) and $|\kappa|\ll$1, $r$ $>$ 0 (corresponding to a structurally-driven transition):
\begin{enumerate}[i)]
	\item for $|r|\ll1$ and $k > 0$:  $X_{phon} \approx - \frac{\lambda}{\kappa} \Delta_1$ and $\Delta_1^2~=~-~\Big(r - \frac{\lambda^2}{\kappa}\Big)/\Big(g + v \frac{\lambda^2}{\kappa^2}\Big)$ ,
	\item for $|\kappa|\ll1$ and $r > 0$:  $\Delta_1 \approx - \frac{\lambda}{r} X_{phon}$ and $X_{phon}^2~=~-~\Big(\kappa - \frac{\lambda^2}{r}\Big)/\Big(u + v \frac{\lambda^2}{r^2}\Big)$.
\end{enumerate}
In both limits the transition occurs at $r = \lambda^2/\kappa$, which defines the new phase transition boundary. In both cases the transition corresponds to a $Z_2$-type of symmetry breaking. \\

\noindent \textbf{B. Phason dispersion}\\

When the excitonic order parameter and the phonon fields are linearly coupled as in equation~(\ref{eq:lang}), there is no rigorous symmetry distinction between different types of symmetry breaking. Furthermore, since the continuous U(1) symmetry is reduced to the $Z_2$ symmetry, there may not be a gapless Goldstone mode \cite{zenker2014fate,murakami2017photoinduced,mazza2019nature}. However, as discussed in detail in Ref.~\cite{Murakami20}, the phason dispersion should give indications of the leading mechanism of instability. In particular, when $\lambda$ is small and the transition is primarily driven by the excitonic part of the model, there are pseudo-Goldstone modes, i.e. modes with a non-zero but small gap that may be understood as arising from the proximity of the U(1) symmetry. This is analogous to chiral quantum chromodynamics, where small mass of $\pi$ mesons can be understood from the perspective of the proximity of the model to the full chiral symmetry \cite{weinberg_95}. From the Lagrangian in equation~(\ref{eq:lang}), the energy of the phason in the symmetry broken phase is given by
\begin{equation}
\omega^2_{phas}(q)  = \frac{\lambda|X_{phon}|}{|\Delta_1|} + \xi q^2.
\end{equation}
In the electronically-driven scenario, $\omega_{phas}^2(q = 0 ) \approx \lambda^2/\kappa$. For small $\lambda$, this is a pseudo-Goldstone mode, i.e. a remnant of the massless Goldstone mode in the U(1) theory that has been gapped by interactions with phonons. On the other hand, in the case of a structurally-driven transition, $\omega_{phas}^2(q = 0 ) \approx r $  and the energy of the phason is as high as that of the amplitude (Higgs) mode.\\

\noindent \textbf{C. Other collective modes}\\

In this paragraph, we compute the dispersion of the remaining Higgs mode and phonon mode, which are now coupled. This is done by expanding the Lagrangian in equation~(\ref{eq:lang}) around the mean-field expectation value, $|\Delta| = \Delta_0 + \delta \Delta$ and $X_{phon} = \expe{X_{phon}} + u $ to quadratic order in the fluctuations. This derivation assumes that the mode energies are less than the quasiparticle gap, so that damping by particle-hole pairs need not be considered.
\begin{widetext}
\begin{equation}
L = \frac{1}{2}\int d^d x \begin{pmatrix} \delta \Delta && u \end{pmatrix} \begin{pmatrix} - \partial_t^2 - r + 3 g \Delta_0^2 + v \expe{X_{phon}}^2 + \xi \partial_x^2 && \lambda + 2 v \expe{X_{phon}} \Delta_0 \\\
\lambda + 2 v \expe{X_{phon}} \Delta_0 &&  - M_{phon} \partial_t^2  - \kappa - 3 u \expe{X_{phon}}^2 - v \Delta_0^2  \end{pmatrix} \begin{pmatrix} \delta \Delta \\ u \end{pmatrix} 
\end{equation}
\end{widetext}
In the frequency-momentum basis, the equations of motion are given by
\begin{widetext}
\begin{equation}
\begin{pmatrix} \omega^2 - r - 3 g \Delta_0^2 - v \expe{X_{phon}}^2 - \xi k^2 && - \lambda - 2 v \expe{X_{phon}} \Delta_0 \\
-\lambda - 2 v \expe{X_{phon}} \Delta_0 &&  - M_{phon} \omega^2  - \kappa - 3 u \expe{X_{phon}}^2 - v \Delta_0^2  \end{pmatrix} \begin{pmatrix} \delta \Delta \\ u \end{pmatrix} = 0
\end{equation}
\end{widetext}
The dispersion relation of the collective modes can be found by solving the secular equation
\begin{widetext}
\begin{subequations}
	\begin{align}
	\left| \begin{pmatrix} \omega^2 - r - 3 g \Delta_0^2 - v \expe{X_{phon}}^2 - \xi k^2 && - \lambda - 2 v \expe{X_{phon}} \Delta_0 \\
	- \lambda - 2 v \expe{X_{phon}} \Delta_0 &&  - M_{phon} \omega^2  - \kappa - 3 u \expe{X_{phon}}^2 - v \Delta_0^2  \end{pmatrix} \right| = 0, \\
	( \omega^2 - C_1(k^2) ) ( \omega^2 - C_2) - C_3 = 0,
	\end{align}
\end{subequations}
\end{widetext}
where $C_1(k^2) = r + 3 g \Delta_0^2 + v \expe{X_{phon}}^2 + \xi q^2 $, $C_2~=~\frac{\kappa + 3 u \expe{X_{phon}}^2 + v \Delta_0^2}{M_{phon}} $ and $C_3 =\frac{\left( \lambda + 2 v \expe{X_{phon}}\right)^2}{M_{phon}} $. The two collective mode energies are expressed as
\begin{equation}
\omega^2_{1,2} = \frac{C_1(q^2) + C_2}{2} \pm \frac{\sqrt{(C_1(q^2) - C_2)^2 - 4 C_3^2 }}{2}.
\end{equation}
Taking the limit of a weakly-coupled system (where $\lambda$ is small but finite), the two modes are $\omega_{1,2}^2~=~C_{1,2}(q^2)$. Approaching the phase boundary either from the electronically-driven direction ($|r|\ll1$, $\kappa > 0$) or the structurally-driven direction ($|\kappa|\ll1$, $r > 0$), we find that there is one mode in each case that goes to zero at the transition, either the amplitude mode or the phonon mode respectively, while the other modes' frequency is hardly affected by the transition. A detailed account of the influence of these modes on the linear and nonlinear optical response is given in Ref. \cite{Murakami20}.

\section*{Supplementary Note 2: Transient electronic distribution and lattice temperature}

\noindent \textbf{A. Transient electronic distribution}\\

In this section, we estimate the electronic temperature ($T_e$) and other properties of the nonequilibrium electron distribution induced by the pump in Ta$_2$NiSe$_5$, following arguments previously given in related contexts \cite{He16,Eckstein11}. Since the crystals have a thickness larger than 1 mm (i.e. orders of magnitude larger than the penetration depth of the pump pulse), we assume that there is no transmission of the pump beam through the sample. Therefore, the pump energy that is not reflected at the sample surface is absorbed. Relying on the optical properties of the material and the experimental parameters, we can estimate the absorbed fluence for every measurement performed. For example, when our pump photon energy is $\Omega_{pump}$~=~1.55~eV and the light beam is polarized parallel to the material's $c$-axis, we have that the reflectance at 11-14 K is $R$ = 0.455 \cite{larkin2018giant, larkin2016excitonic}. Thus, for an incident fluence of 1.56 mJ/cm$^{2}$, the corresponding absorbed fluence is 0.85 mJ/cm$^2$. We can estimate the energy density absorbed to be $E_{v}$ = 65~meV per formula unit. Since each absorbed photon creates a particle-hole pair at energy 1.55~eV, we estimate that the pump produces a density of $n_{pump}\approx$ 0.04 pairs per formula unit. At the  minimum fluence used in our trARPES experiments the absorbed fluence is $\sim$0.4 mJ/cm$^2$, corresponding to an energy of about 30 meV per formula unit, and for the UED experiments the absorbed fluence used is $\sim$0.1 mJ/cm$^2$ corresponding to an energy of about 7.6 meV per formula unit.

The initial dynamics is entirely within the electronic subsystem, whose energy is thus conserved. Solution of the quantum Boltzmann equation in related contexts \cite{He16} indicates that the distribution thermalizes on a very short (few fs) timescale while conserving the number of electron-hole pairs. Then, over a longer (10-50~fs) timescale, an inverse Auger process \cite{Eckstein11} downscatters a carrier from the high energy tail of the distribution thereby lowering the temperature  while creating a particle-hole pair. The result of this process is an intermediate-time distribution characterized by a  lower temperature and a higher density of particle-hole pairs, both determined by balancing Auger and inverse Auger processes. This distribution then loses energy to the lattice on phonon-timescales and the density also recombines at a yet slower rate both from Auger and multiphonon processes. We may estimate the density of pairs in the intermediate time state by equating the Auger up and down scattering and assuming that the energy is the same as the initial energy.  Following Ref.~\cite{He16}, by making a dynamical mean-field-like approximation in which the momentum dependence of the scattering rate is averaged over, and taking the CB and VB densities of states to be the same so the electrons and holes have the same distribution functions and  in steady-state the same chemical potential.  The amplitude for the Auger upscattering (one particle at high energy $E_1+E_2+E_3)$ to one particle $E_1$  plus an across bandgap excitation of one particle at $E_2$ and one at $E_3$)  is then proportional to $f(E_1+E_2+E_3)(1-f(E_1))(1-f(E_2))(1-f(E_3))$ where $f$ is the distribution function, while the downscattering is proportional to $f(E_1)f(E_2)f(E_3)(1-f(E_1+E_2+E_3))$. 
Equating the up and down scattering gives
\begin{widetext}
\begin{eqnarray}
&&\int (dk_1)(dk_2)(dk_3)\left[f(E_1)f(E_2)f(E_3)(1-f(E_1+E_2+E_3))\right]=
\label{steadystate}
\\&&\int  (dk_1)(dk_2)(dk_3)\left[f(E_1+E_2+E_3)(1-f(E_1))(1-f(E_2))(1-f(E_3))\right]
\nonumber
\end{eqnarray}
\end{widetext}
where $(dk)$ is the appropriate integration measure in momentum space. We assume that the distributions are Fermi functions and that, in the presence of the gap $\Delta$, the energy dispersion is $E=\sqrt{\varepsilon_k^2+\Delta^2}$, and the density of states $\nu(\varepsilon)=\int(dk)\delta(\varepsilon-\varepsilon_k)$ is constant and identical for the CB and VB. We define the  fugacity $\zeta=e^\frac{\mu-\Delta}{T}$, find the $\zeta(\beta)$ that yields the desired energy density $\mathcal{E}$ assuming $\mathcal{E}=2\int(dk)Ef(E_k)$ (the $2$ is from summing over electrons and holes)  and then find the $\beta$ that solves Eq.~\ref{steadystate}. Using $\Delta$ = 0.15 eV and a density of states per formula unit $\nu=2/eV$, we find the results presented in Table ~\ref{Occupancytab}.\\

\begin{table}[tb]
	\label{Occupancytab}
	\begin{tabular}{ccccc}
		\hline
		F (mJ/cm$^2$) & E (meV/fu) & $n_{ss}$ & T (K) & VB edge occ 
		\\
		\hline\hline
		0.85 (trARPES, $F_{max}$) & 65 & 0.12 & 1600 & 0.75 
		\\
		0.40 (trARPES, $F_{min}$) & 30 & 0.07 & 1200 & 0.82 
		\\
		0.10 (UED) & 7.6 & 0.020 & 700 & 0.92
		\\
		\hline
	\end{tabular} 
	\caption{Energy, number of excited electrons per formula unit,  temperature, and occupancy of state at VB edge for pseudoequilbrium distribution generated by electron equilibration at fluences used in experiment}
	\label{Occupancytab}
\end{table}

\noindent \textbf{B. Transient lattice temperature}\\

In this paragraph, we estimate the maximum transient lattice temperature that can be reached in Ta$_2$NiSe$_5$ after photoexcitation. We rely on the simple thermodynamic expression
\begin{equation}
Q = \int_{T_{L,in}}^{T_{L, fin}} mC(T) dT
\end{equation}
where $Q$ is the absorbed heat from a single laser pulse, $m$ is the illuminated mass, $C(T)$ is the temperature-dependent specific heat, $T_{L, in}$ is the initial equilibrium temperature and $T_{L, fin}$ is the final lattice temperature. We calculate $m$ through the material density $\rho$~=~7.72~g/cm$^3$ and the illuminated sample volume $V$. Finally, we rely on the temperature dependence of the heat capacity, as measured in Ref. \cite{lu2017zero}. For the trARPES experiments (where $T_{L, in}$ = 14 K), at the largest absorbed fluence of 0.85 mJ/cm$^2$ the calculation yields a maximum $T_{L, fin}$ = 210 K, which lies well below $T_C$~=~328~K. For the UED experiments (where $T_{L, in}$ is either 77 K or 295 K), the calculation yields a maximum $T_{L, fin}$ of 89 K or 302 K for the absorbed fluence of 0.1~mJ/cm$^2$. Both values lie again below $T_C$.

\begin{figure}[b]
	\begin{center}
		\includegraphics[width=0.8\columnwidth]{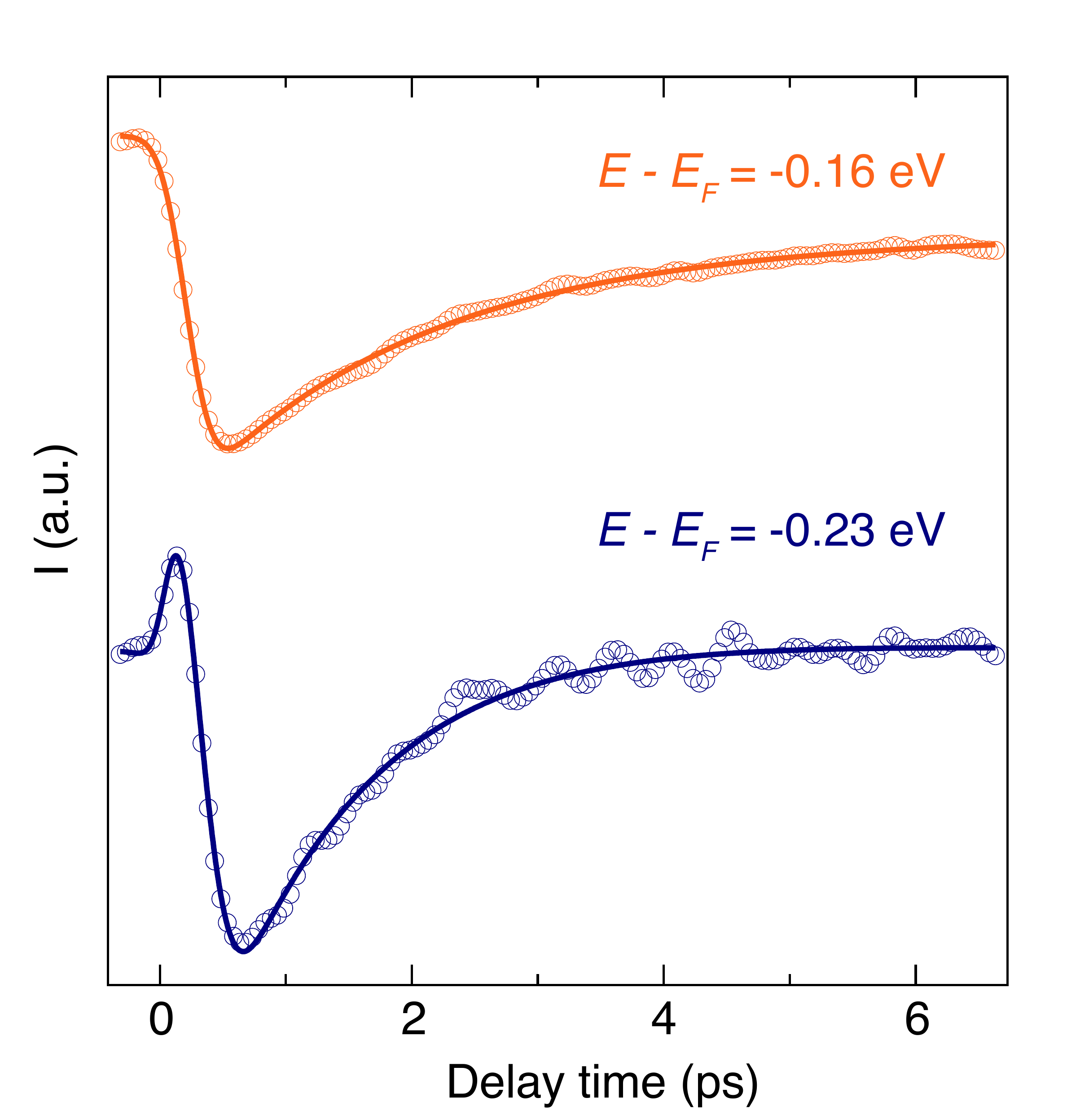}
	\end{center}
	\caption{\textbf{Temporal resolution in the Ti:Sapphire-based trARPES measurement.} Temporal traces selected at the energies of -0.16~eV (orange dots) and -0.23~eV (blue dots), referenced to $E_F$. The solids lines are fit to the data. The overall time resolution can be estimated $160\pm30$\,fs, corresponding to the first rise of the trace at -0.23 eV. The data have been acquired at 14 K using the 6.20 eV probe trARPES scheme and an absorbed fluence of 0.85 mJ/cm$^2$.}
	\label{fig:TimeResolution}
\end{figure}

\section*{Supplementary Note 3: Time resolution in the experimental setups}

Here, we estimate the time resolution of our trARPES and UED setups. For trARPES, Fig.~\ref{fig:TimeResolution} shows two temporal traces at energies of -0.16~eV (orange dots) and -0.23~eV (blue dots), referenced to $E_F$. The time resolution is extracted from the fast rise appearing in the temporal trace at -0.23~eV, which is assumed to be time-resolution limited and therefore follows the cross-correlation between the pump and the probe pulses. Specifically, the value of $160\pm30$\,fs is obtained by fitting the derivative of the trace at -0.23~eV around zero time delay with a Gaussian function; the temporal resolution is taken as its full-width-at-half-maximum (FWHM). Given this instrumental resolution, the rise time of the trace at -0.16~eV is therefore not resolution limited and involves a phenomenon rising with a $\sim$150 fs timescale. As such, the temporal trace reaches its minimum value around 0.3-0.4 ps. This slow response is consistent with that reported in previous trARPES transient reflectivity experiments \cite{mor2017ultrafast,werdehausen2016coherent,mor_inhibition,werdehausen2018photo,tang2020non}.

\begin{figure}[t]
	\centering
	\includegraphics[width=0.8\columnwidth]{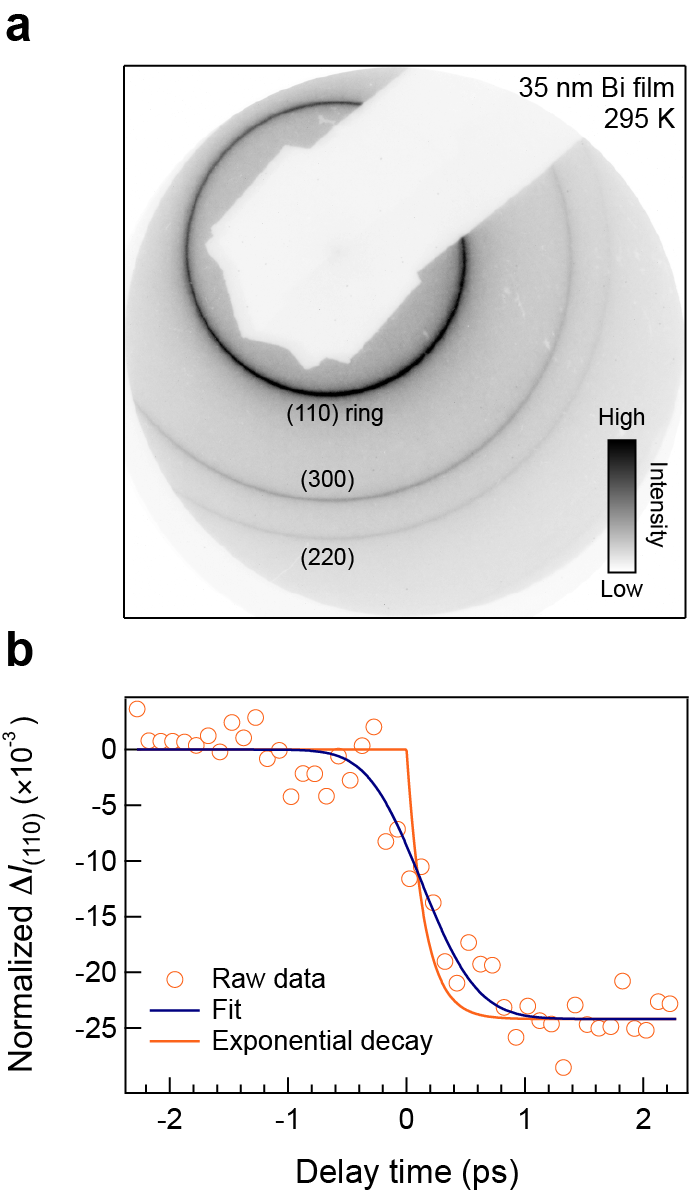}
	\caption{\textbf{Temporal resolution in the UED measurement.} \textbf{a,}~Static diffraction pattern at room temperature of a 35\,nm Bi film, which was deposited on a 10\,nm-thick silicon nitride window via electron beam evaporation. The three most prominent Debye-Scherrer rings are labeled. \textbf{b,}~Photoinduced change in the (110) ring intensity, normalized to its value before photoexcitation. Data was taken at 295\,K, with a repetition rate of 1\,kHz and incident pump fluence of 0.54\,mJ/cm$^2$. Following Ref.~\cite{weathersby2015mev}, the instrumental temporal resolution is determined by fitting the raw data (red circles) with an exponential decaying function (time constant 150\,fs, red curve) convolved by a Gaussian kernel. The temporal resolution, $0.8\pm0.1$\,ps, is taken as the fitted FWHM of the Gaussian function.}
	\label{fig:ued-bi}
\end{figure}

For UED, we determine the temporal resolution by using a Bi film whose response to photoexcitation is known and fast \cite{weathersby2015mev}. Tracking the evolution of the most intense Debye-Scherrer ring (Fig.\,\ref{fig:ued-bi}a), we model the fast lattice response by an exponential function convolved with a Gaussian kernel (Fig.~\ref{fig:ued-bi}b). The temporal resolution of 0.8~$\pm$~0.1\,ps is taken as the fitted FWHM of the Gaussian function.

All time traces of our trARPES and UED experiments are fit with a model function composed of two exponential functions. The first captures the fast initial response and the second describes the slower relaxation dynamics. Specifically, the function is
\begin{eqnarray}\nonumber
f(t) &=& \Big\{\Theta(t-t_0)\cdot\left[A_1\left(1 - e^{-(t-t_0)/\tau_1}\right) +\right.\\
&& \left. A_2\left(1 - e^{-(t-t_0)/\tau_2}\right)\right]\Big\}\ast \mathcal{G}(w),
\end{eqnarray}
where $t$ is the pump-probe delay time, $t_0$ is the time zero when the pump and probe pulses maximally overlap in time, $\Theta(\cdot)$ is the Heaviside step function, $\ast$ denotes convolution, and $\mathcal{G}(w)$ is a normalized Gaussian function with FWHM $w$, where $w$ is the instrumental temporal resolution. The fitting parameters are $t_0$, $A_{1,2}$, and $\tau_{1,2}$.

\section*{Supplementary Note 4: Energy-momentum dispersion}

\begin{figure}[b]
	\begin{center}
		\includegraphics[width=\columnwidth]{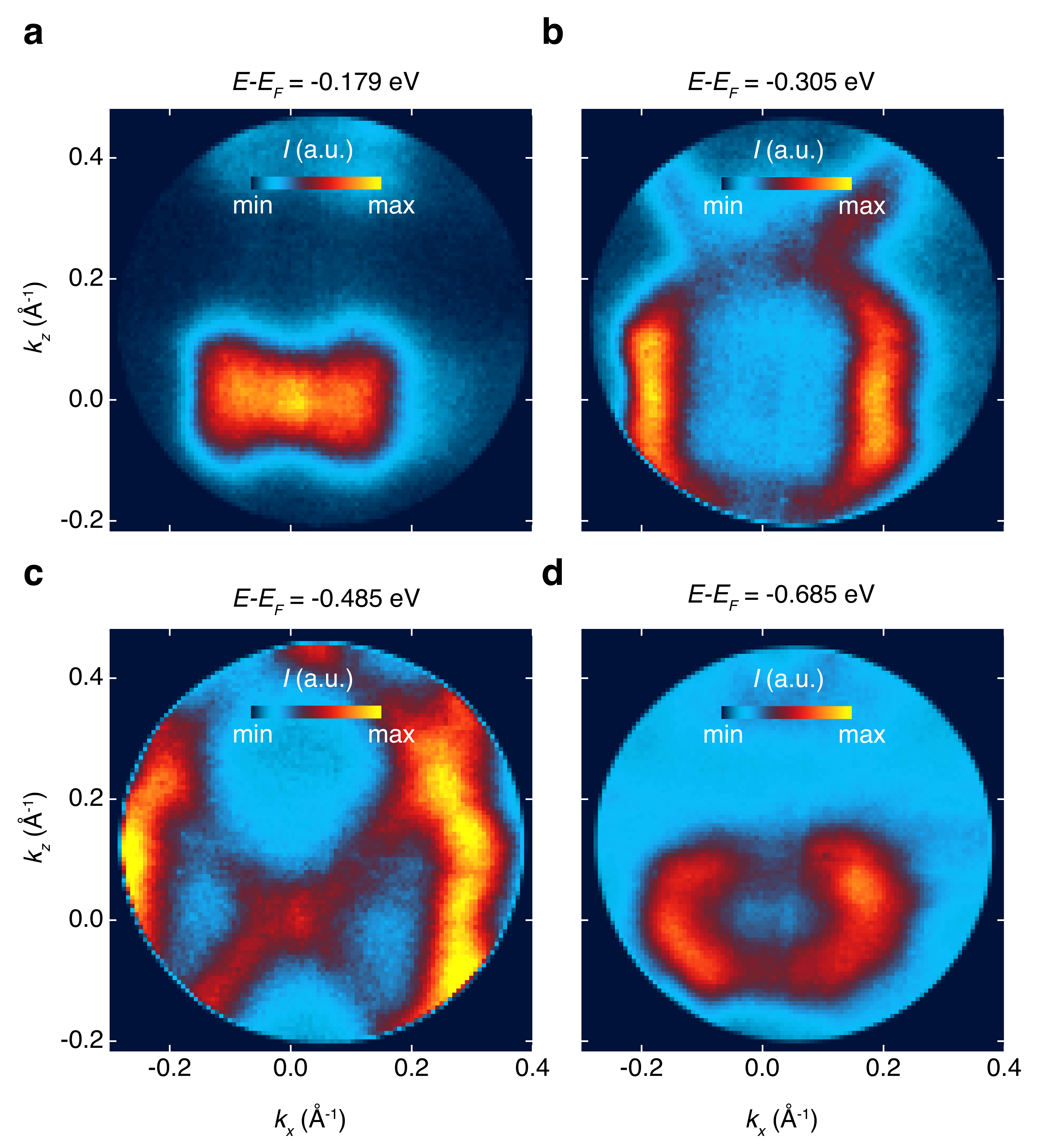}
		\caption{\textbf{Constant-energy trARPES maps.} Constant-energy cuts of the trARPES data at 11 K in the $k_x$-$k_z$ momentum space, measured before photoexcitation ($t$ $<$ 0) with the 10.75 eV probe scheme. Representative energies are shown and referenced to $E_F$: \textbf{a,}~-0.179 eV; \textbf{b,}~-0.305 eV; \textbf{c,}~-0.485~eV; \textbf{d,}~-0.685 eV.}
		\label{fig:HolePocket}
	\end{center}
\end{figure}

\begin{figure*}[t]
	\begin{center}
		\includegraphics[width=2\columnwidth]{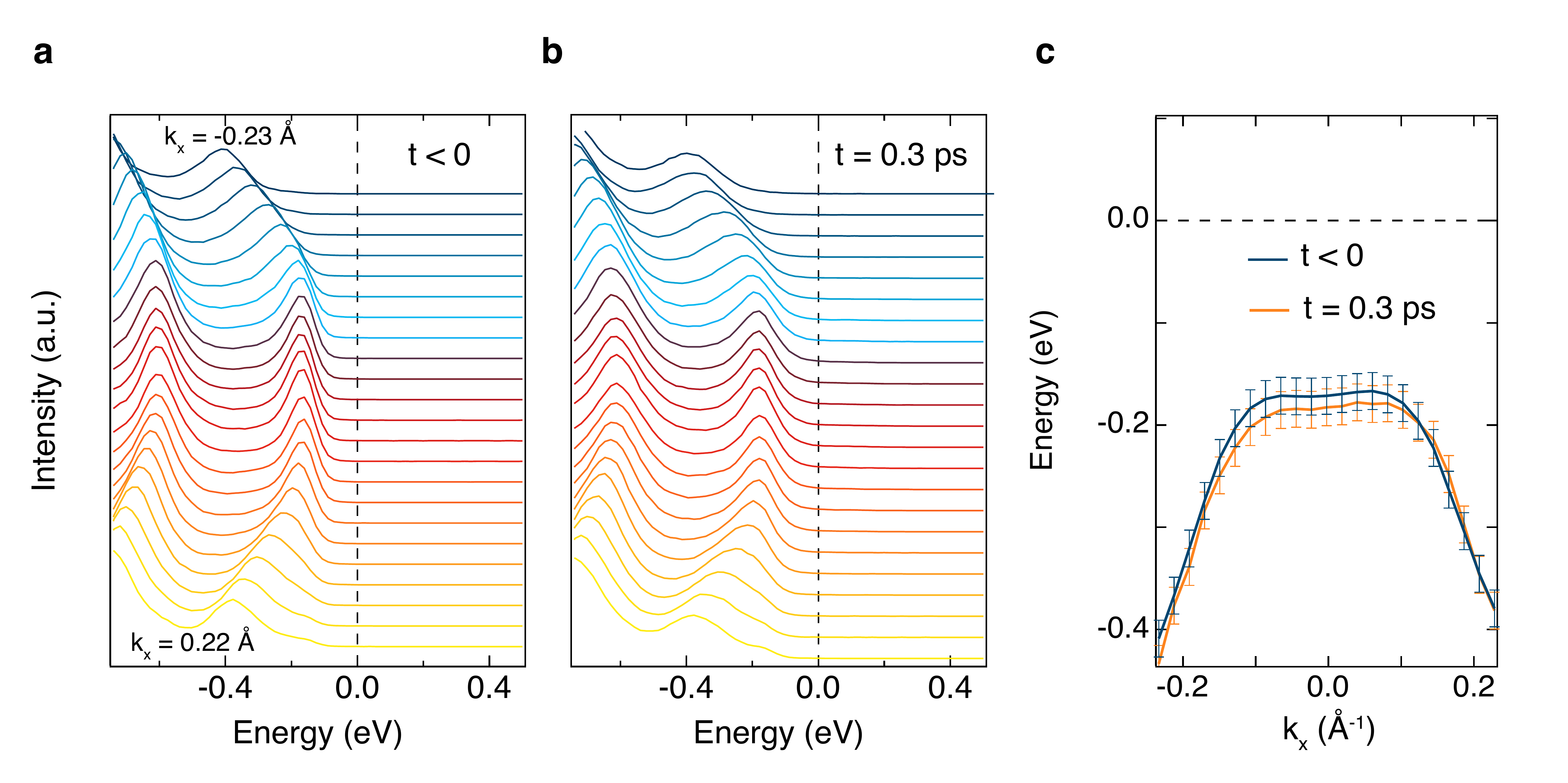}
		\caption{\textbf{Energy-momentum dispersion relation.} \textbf{a,b}, EDCs along $k_x$, as obtained from the spectra in Figs. 2a,b before time zero and at $t$ = 0.3 ps. \textbf{c}, Energy-momentum dispersion relation before time zero and at the maximum signal response, as obtained from tracking the VB peak energy at different momenta around the center of the Brillouin zone. The VB peak energy at different momenta has been located by finding the maximum intensity. The error bars have been estimated accounting for the energy spacing between two adjacent data points.}
		\label{fig:Dispersion}
	\end{center}
\end{figure*}

In this section, we focus on the electronic structure measured in our trARPES experiments. Figure \ref{fig:HolePocket} shows representative constant-energy cuts of the data acquired before photoexcitation ($t$ $<$ 0) at 11 K with a probe photon energy of 10.75 eV. These data show the VB structure in the $k_x$-$k_z$ momentum space, where $k_{x}$ denotes the direction parallel to the Ta and Ni chains in the orthorhombic cell and $k_{z}$ the one perpendicular to the chains. The energy, referenced to $E_F$, at which each constant-energy map is cut is indicated in the labels. In particular, Fig. \ref{fig:HolePocket}a corresponds to the upper part of the VB and exhibits an anisotropic hole-like pocket centered around the $\Gamma$ point of the Brillouin zone. Increasing the binding energy (Fig. \ref{fig:HolePocket}b-d) gives access to a rich pattern of bands, in agreement with previous static ARPES data \cite{wakisaka2012photoemission,watson2019band}. Therefore, our results in equilibrium are consistent with the band structure known from the literature of Ta$_2$NiSe$_5$. Similar results were obtained in the measurements with the 6.20~eV probe scheme, albeit over the narrower $k_x$-$k_z$ range expected from this photon energy.

Figure \ref{fig:Dispersion} shows the energy distribution curves (EDCs) along $k_x$, as obtained from the spectra in Fig. 2a,b before time zero and at the maximum signal response. As discussed in the main text, we observe that the dominant effect of photoexcitation on the VB is the modification of its intensity and width at all momenta. On the other hand, its peak position is barely affected by photoexcitation. Figure \ref{fig:Dispersion}c shows the energy-momentum dispersion before time zero and around the maximum signal response, as obtained from tracking the VB peak energy at different momenta in the explored range. We observe that at this time delay the VB peak is shifted toward higher binding energy in the momentum region where the VB is flat, whereas no sizeable changes are detected on the wings of the band. This momentum dependence demonstrates that the detected shifts are not caused by a change in the workfunction of Ta$_2$NiSe$_5$. In Supplementary Note~5C, we show how the intensity, width, and peak position of the VB vary as a function of time.

\begin{figure*}
	\begin{center}
		\includegraphics[width=1.5\columnwidth]{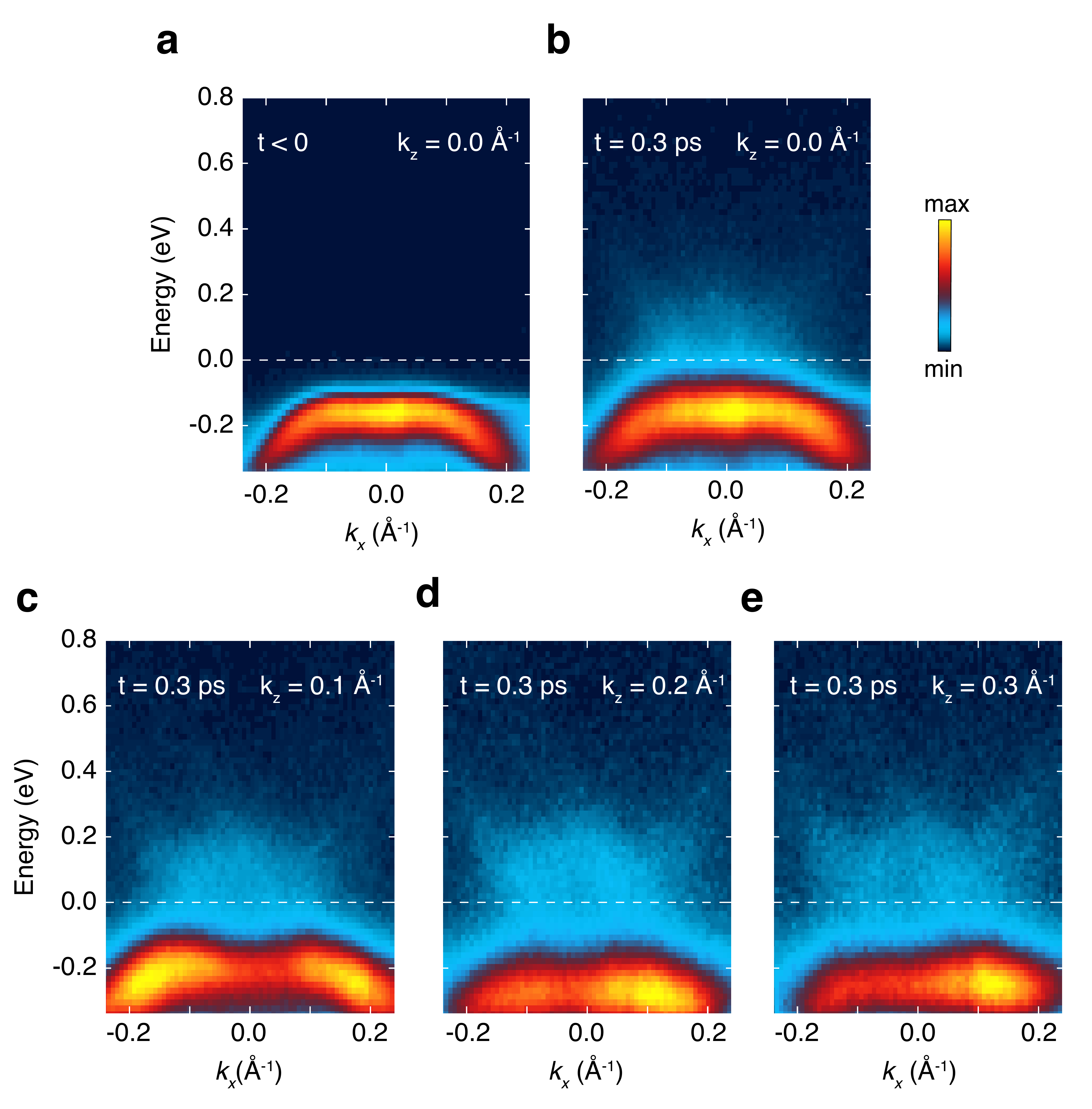}
		\caption{\textbf{Analysis of the conduction band.} Snapshots of the trARPES spectra along the $k_x$ momentum direction. \textbf{a},~Spectrum at $t$ $<$ 0 and $k_z$ = 0.0 \AA$^{-1}$. \textbf{b-e}, Spectra at $t$ = 0.3 ps and for representative $k_z$ momenta, as indicated in the labels. The spectral weight above $E_F$ assumes a W-like shape consistent with the dispersion of the CB. The VB retains its flattened M-like shape. Both features indicate that the hybridization gap is still robust and can be estimated in a range between 180 and 210 meV at $k_z$ = 0.0~\AA$^{-1}$. The snapshots at finite $k_z$ demonstrate that the VB and CB never cross each other in the whole $k_x$-$k_z$ momentum space around $\Gamma$ and thus the electronic gap size remains finite. The data have been measured at 11 K with a probe photon energy of 10.75 eV and an absorbed pump fluence of 0.4 mJ/cm$^2$.}
		\thispagestyle{empty}
		\label{fig:CB}
	\end{center}
\end{figure*}

Finally, in Fig. \ref{fig:CB} we display trARPES snapshots obtained with improved sensitivity around $E_F$ at 11 K. The snapshot in panel (a) shows the data acquired before photoexcitation ($t$ $<$ 0) and $k_z$ = 0.0 \AA$^{-1}$, whereas panels (b-e) represent an enlarged version of Fig. 3c-f and are obtained at $t$ = 0.3 ps after photoexcitation with an absorbed pump fluence of 0.4 mJ/cm$^2$. As underlined in the main text, in Fig. \ref{fig:CB}a we observe only the flattened VB around the $\Gamma$ point of the Brillouin zone. Pumping out of equilibrium (Fig. \ref{fig:CB}b-e) leads to the appearance of an upward-dispersing CB with the characteristic W-like shape expected from band hybridization and our \textit{ab initio} calculations in the monoclinic phase (see Fig. 1b and Supplementary Note 7). As mentioned above, the VB undergoes depletion and broadening, but no distortions of its shape are found compared to the spectra before photoexcitation. This is unlike the equilibrium case above $T_C$, at which the gap collapses and the two bands intersect, losing their hybridized M-like and W-like shapes \cite{watson2019band}. The bottom of the CB lies around 50 meV above the chemical potential, indicating that the latter is pinned closer to the CB minimum than the VB maximum. This aspect is consistent with the conclusions of scanning tunneling spectroscopy (STS) data \cite{lee2019strong}, although the size of the electronic gap in that equilibrium study (300 meV at 78 K) is larger than the electronic gap found by us and the optical gap measured by spectroscopic ellipsometry ($\sim$200 meV at 10 K) \cite{lu2017zero,larkin2018giant}. In Supplementary Note~7D we explain this discrepancy quantitatively by noting that the STS data have been obtained on a sample with a monoclinic angle $\beta$ = 92.5$^\circ$ that is much larger than the one expected in the monoclinic phase (90.5$^\circ$-90.6$^\circ$) \cite{sunshine1985structure,nakano2018antiferroelectric}. As such, the STS gap increases compared to the gap of a crystal with the nominal $\beta$ angle. In our trARPES data, the uncertainty provided by the energy resolution and the simultaneous presence of a photoinduced carrier density does not allow us to extrapolate an absolute value for the gap size in equilibrium conditions. We can only establish that, close to the $\Gamma$ point of the Brillouin zone ($k_x$~$\sim$~$\pm$0.05~\AA$^{-1}$, $k_z$~$\sim$~$\pm$0.00~\AA$^{-1}$), the gap size lies in a range between 180~meV and 210 meV even if $T_e$ is well above $T_C$ by several hundred kelvins. This observation allows us to exclude the scenario in which the gap opening in Ta$_{2}$NiSe$_{5}$ has a dominant (or even substantial) EI nature. The consequence of this observation on the fate of the EI order parameter and its collective modes is explained in Supplementary Note 1.

\section*{Supplementary Note 5: Origin of the rise time in the trARPES data}

In order to determine the origin of the rise time in the trARPES data of Fig. 3a-c, we examined several physical processes. Relying on this analysis and on the direct observation of phonon oscillations on the rise of the response (Fig. 3d and Fig. \ref{fig:coherent}), we conclude that the rise time is phononic in nature.\\

\noindent \textbf{A. Screening of the Coulomb interaction}\\

First, we consider a scenario in which the photoexcited carrier density creates a transient metallic state that screens the electron-hole Coulomb interaction. This process is present in all materials and lies at the origin of the so-called bandgap renormalization observed in standard band semiconductors \cite{beni1978theory}. In the case of a pure EI (defined as a material whose gap is solely driven by an excitonic instability), this mechanism acquires special importance: indeed, increasing $T_e$ $\gg$ $T_C$ is expected to cause a collapse of the excitonic gap. In the following, we investigate this scenario in detail relying on our band structure calculations (Supplementary Note 7). We determine the plasma frequency $\omega_{p}$ (i.e. the relevant parameter governing the screening dynamics) along the three crystallographic directions as a function of the Fermi energy. We use the standard procedure of averaging the squared Fermi velocity over the 
Fermi surface. The result is then convoluted with the Fermi function corresponding to the $T_e$ reached in our trARPES experiment (see Supplementary Note 2). We obtain that $\omega_{p,a}$~=~2.07~eV, $\omega_{p,b}$~=~0.28~eV, and $\omega_{p,c}$~=~0.59~eV, along the $a$, $b$, and $c$ crystallographic axes, respectively. These values imply that the screening time $\tau_{scr}$ is on the order of 2-15~fs, which is significantly faster than the timescale discussed in this paper.

Another way to estimate $\tau_{scr}$ is through the simplified formula \cite{hellmann2012time}
\begin{equation}
\tau_{scr} =  2\pi\sqrt{\frac{\epsilon_0\epsilon_{opt}}{e^2}\frac{m^*}{\eta n_{ph}}},
\end{equation}
where $\eta$ is the quantum efficiency, $n_{ph}$ is the photoexcited carrier density, $m^*$ is the carrier effective mass, and $\epsilon_{opt}$ is a non-resonant dielectric constant that accounts for the background polarizability of the valence electrons (i.e. the real part of the dielectric function at optical frequencies). Substituting the relevant parameters for Ta$_2$NiSe$_5$ \cite{larkin2018giant} and assuming that $\eta$ = 1 yields a fluence-dependent rise time of the order of $\tau_{scr}$~$\sim$~10 fs. These results are very robust with respect to the choice of the effective dielectric constant accounting for the material's anisotropy and the carriers' effective mass. Therefore, the estimate is in agreement with the more refined \textit{ab initio} calculation given above and confirm that screening of the Coulomb interaction (and bandgap renormalization due to an increase of $T_e$) cannot explain the dominant response observed in our trARPES data. \\

\noindent \textbf{B. Quasiparticle avalanche multiplication}\\

Since a putative EI bears similarities with a superconductor, we also consider the processes typically observed in the nonequilibrium response of superconductors photoexcited with photon energies much larger than the gap size. The generally-accepted picture for describing the initial ultrafast dynamics of a superconductor involves the process of quasiparticle avalanche multiplication \cite{han1990femtosecond,kusar2008controlled,Stojchevska}. In this mechanism, a photon with energy exceeding the superconducting gap creates a particle-hole pair in the material. This particle-hole pair then loses its energy very rapidly through avalanche scattering with high-energy bosons within 0.1~ps, creating a large nonequilibrium boson population in the process. A subset of bosons has an energy exceeding the superconducting gap and the condensate is depleted by absorbing energy from the hot bosonic bath. In most superconductors, the bosons have been identified as high-energy phonons, but contributions from spin excitations have also been proposed \cite{dal2015snapshots}. The number of quasiparticles created per absorbed photon can be approximately evaluated by the ratio between the initial photon energy and the superconducting gap \cite{han1990femtosecond}. The quasiparticles recombine again and a bottleneck establishes, with the high-energy bosons being in quasi-equilibrium with the quasiparticles. In certain parameter regimes, this phenomenon can lead to the depletion of the superconducting condensate within 0.5-1~ps, causing the slow rise of the (optical or photoemission) pump-probe response \cite{kusar2008controlled, Stojchevska, boschini2018collapse}. Together with the high-energy bosons, in the avalanche multiplication also low-energy bosons are emitted. However, these do not have enough energy to contribute to the condensate vaporization. 

In order for the quasiparticle avalanche multiplication to govern the rise-time of the pump-probe response, we have to assume that, after photoexcitation, hot phonons are emitted with an energy exceeding the size of the single-particle gap. In Ta$_2$NiSe$_5$, the $\sim$180-300 meV gap \cite{lu2017zero,larkin2018giant,lee2019strong} is too large to account for a quasiparticle avalanche multiplication process, as the maximum phonon energy at the $\Gamma$ point of the Brillouin zone is $\sim$37 meV \cite{larkin2018infrared,werdehausen2016coherent}. Therefore, it is highly unlikely that multi-phonon excitation processes from the photogenerated electron-hole pairs can be responsible for the VB delayed response. Moreover, the direct visualization of the low-energy dynamics in the VB supports this interpretation, as no significant variation of the VB response time with increasing fluence is observed in Ta$_2$NiSe$_5$.\\

\noindent \textbf{C. Phonon occupation}\\

\begin{figure*}
	\begin{center}
		\includegraphics[width=1.5\columnwidth]{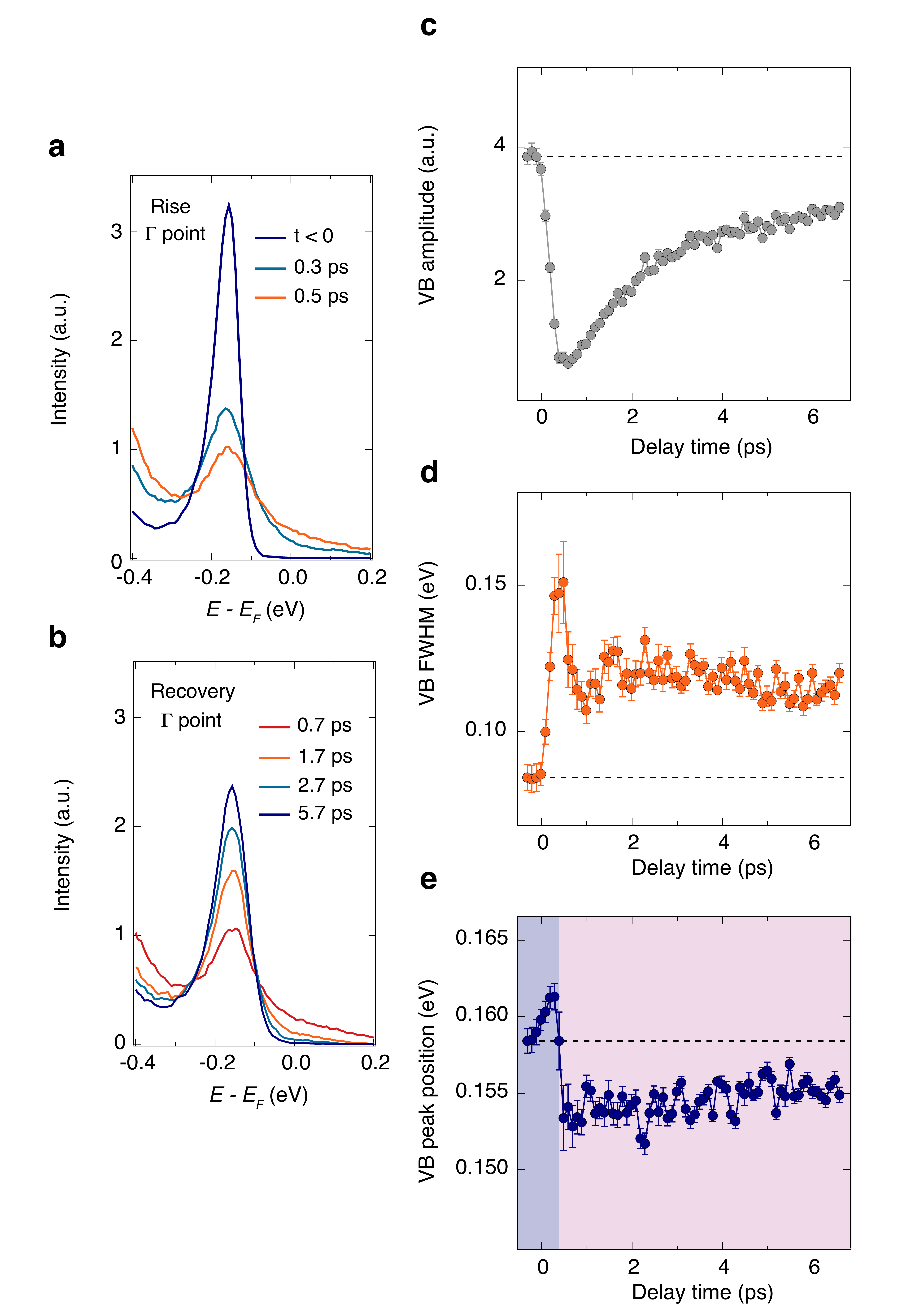}
		\caption{\textbf{Ultrafast dynamics of the VB.} Time evolution of the EDCs at the $\Gamma$ point of the Brillouin zone during the \textbf{a},~rise and \textbf{b}, recovery of the response. The data are measured at 14 K with 6.20~eV probe photon energy and an absorbed pump fluence of 0.85 mJ/cm$^2$. Ultrafast dynamics of the \textbf{c}, VB amplitude; \textbf{d}, VB FWHM; and \textbf{e}, VB peak position. All parameters are extracted from a Lorentz fit to the EDCs.}
		\label{fig:Lorentz}
		\thispagestyle{empty}
	\end{center}
\end{figure*}

\begin{figure}
	\centering
	\includegraphics[width=0.85\columnwidth]{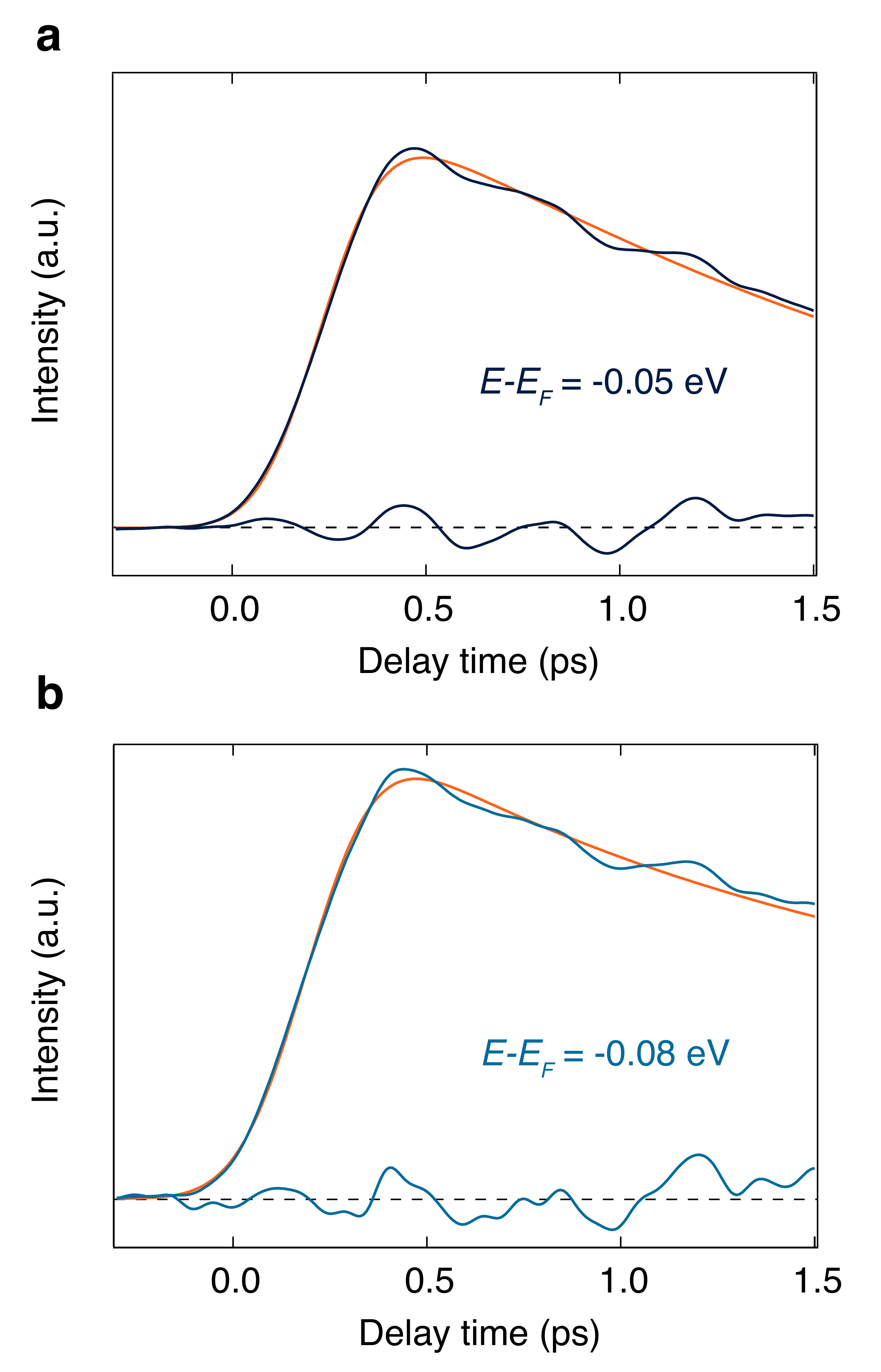}
	\caption{\textbf{Coherent phonons on the rise of the response.} Time dependence of the momentum-integrated photoelectron intensity around an  energy of \textbf{a,} -0.05 eV and  \textbf{b}, -0.08 eV, referenced to $E_F$. The energy interval over which the intensity is integrated around the specified  energy is $\pm$50 meV. The residual from a multiexponential fit to the data is shown in each panel. The fit function is displayed as an orange curve. For visualization purposes, the data have been smoothed and the residuals have been multiplied by a factor of 2.}
	\label{fig:coherent}
\end{figure}

In this section, we correlate the slow rise time in the VB response to the buildup of the phonon occupation. Specifically, we disentangle the contributions of the electronic and phononic degrees of freedom to the VB renormalization and show that phonons play the most significant role in the observed dynamics. To this aim, we extract quantitative information on the time evolution of the VB response by fitting the EDCs with a phenomenological model based on a Lorentzian function and a background. To maintain high time resolution, we focus on the data measured with the 6.20~eV probe beam, shown in Fig. \ref{fig:Lorentz}a,b for representative time delays. The time evolution of the relevant VB parameters (amplitude, FWHM, and peak position) extracted from the Lorentz fit are shown in Fig. \ref{fig:Lorentz}c-e. We observe that the VB amplitude (Fig.~\ref{fig:Lorentz}c) drops upon photoexcitation with a finite rise time, reaching its minimum value around 0.4~ps and recovering with an biexponential behavior (with timescales of $\sim$1.7 ps and $\sim$20 ps). The amplitude signal measures the depletion dynamics of the VB after the arrival of the pump pulse. The delayed rise and subsequent recovery indicates that the photoexcited holes (which are initially created with a high excess energy inside the VB) cool down toward the VB edge in 0.4~ps through the emission of optical phonons and recombine following two distinct decay processes. Therefore, we attribute the timescale of 0.4~ps to the optical phonon-mediated intraband cooling dynamics of the holes. The time evolution of the VB FWHM (Fig. \ref{fig:Lorentz}d) quantifies the change of the imaginary part of the self-energy in the single-particle spectral function describing the system. Upon photoexcitation, the VB broadens almost by a factor of 2 within 0.4~ps and then relaxes to a rather constant value for tens of picoseconds. The initial rise can be ascribed to the increase in the scattering processes with the optical phonons emitted in the cooling dynamics, whereas the constant plateau is reached when acoustic phonons start to dominate the scattering processes (with the carriers lying close to the band edges and undergoing recombination). This picture is consistent with the interpretation provided to the dynamics of the VB amplitude (Fig.~\ref{fig:Lorentz}c). Finally, Fig. \ref{fig:Lorentz}e shows the time evolution of the VB peak energy. Contrary to the response of the VB amplitude and FWHM, the VB peak energy responds promptly to photoexcitation with a small increase of $\sim$3~meV that is complete around 0.1 ps (blue shaded area in the graph). This change is followed by a decrease of $\sim$8 meV, which is maximum around 0.4~ps (red shaded area in the graph). The subsequent dynamics is characterized by a long recovery. Note that the first increase is not an artifact of the fit, since it mimics the resolution-limited rise found in our raw data (see Fig. \ref{fig:TimeResolution}). We attribute this rise to the process of bandgap renormalization driven by the increase of $T_e$ (i.e. in the electronic screening described in Supplementary Note 5A), a process that is immediate with photoexcitation \cite{hein2016momentum, zhu}. As mentioned above, while this process is present in all semiconductors \cite{beni1978theory}, it acquires particular relevance in the case of a pure EI (defined as a material in which the entire hybridization gap is excitonic in nature). In such a pure EI, under our experimental conditions, the gap would undergo complete closure within the electronic timescale set by this screening process. In contrast with this scenario, we observe that the electronic screening in Ta$_2$NiSe$_5$ is small even when $T_e \gg T_C$ (as shown in Fig.~2a,b) and for all time delays during which the carriers persist in the bands. This finding is consistent with the results of our first-principles calculations in Supplementary Section 7D. Therefore, the fact that the gap size remains finite at all time delays even if $T_e \gg T_C$ indicates a significant contribution of the structural degrees of freedom behind gap formation in Ta$_2$NiSe$_5$. Along this line, we ascribe the second effect contributing to the gap variation (i.e. the one that is complete within 0.4~ps) to the buildup of the maximum occupation for the strongly-coupled optical phonons participating in the monoclinic distortion (which is still robust under our experimental conditions, as illustrated in Supplementary Note 7B). Also this scenario finds agreement with our description of the VB amplitude (Fig. \ref{fig:Lorentz}c) and FWHM (Fig. \ref{fig:Lorentz}d) dynamics.

\begin{figure}[b]
	\centering
	\includegraphics[width=0.85\columnwidth]{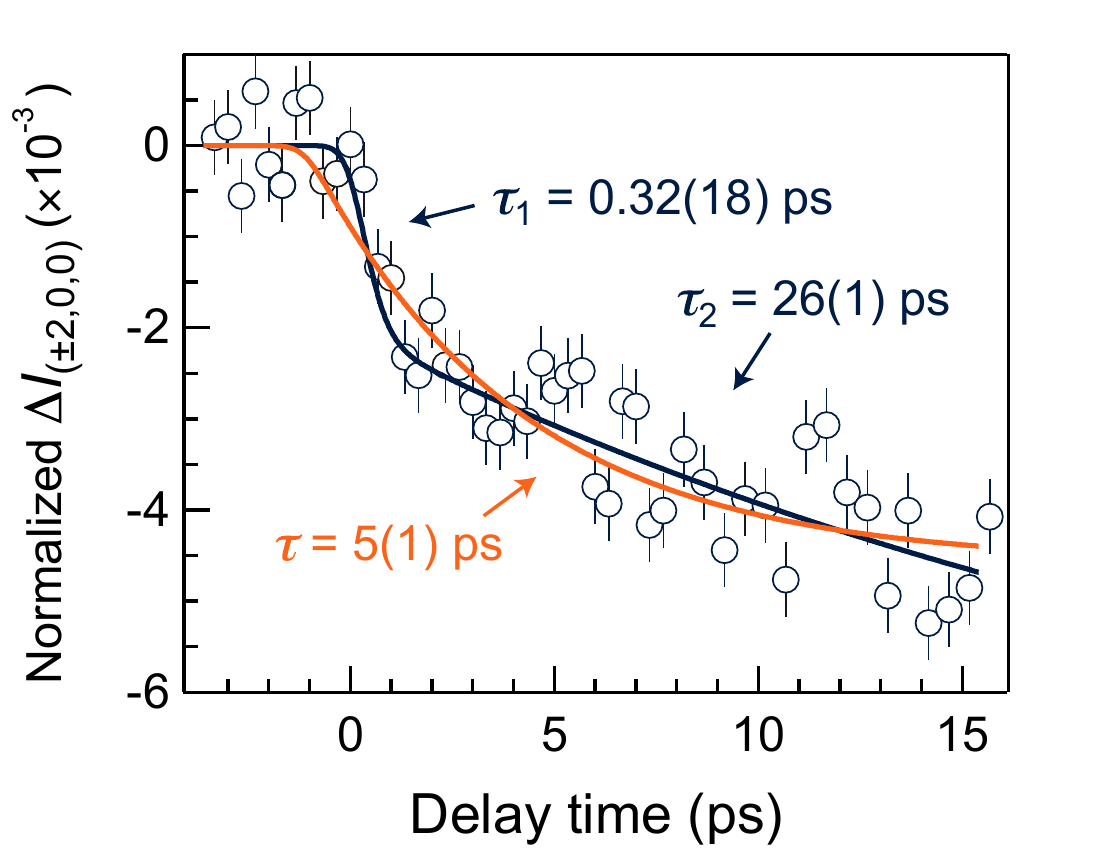}
	\caption{\textbf{Fit to the UED temporal response.} Photoinduced change of the integrated diffraction intensity at room temperature, normalized to its value before excitation. The absorbed fluence is 0.1 mJ/cm$^2$, corresponding to $T_e$ $\sim$ 380 K. Intensity values are taken as the average between (200) and $(\bar{2}00)$ peaks. Fit to the temporal trace using a one-exponential decay (orange line) and a two-exponential decay (blue line) functions, both convoluted with a Gaussian function that accounts for the time resolution of the experiment. The extracted timescales are indicated on the figure. Only the use of a two-exponential decay function captures the correct zero time delay of the response and the fast initial decay.}
	\label{fig:UED_exp}
\end{figure}

\section*{Supplementary Note 6: Analysis of the collective modes in the trARPES data}

In this section, we discuss the features in our trARPES data that indicate that in photoexcited Ta$_2$NiSe$_5$ the crystal remains in the same ground state structure at all time delays. We focus on the four frequency components that appear in the Fourier transform of Fig. 3e. These correspond to Raman-active phonons that are excited coherently via an impulsive/displacive-type generation mechanism \cite{stevens2002coherent}. These modes have been studied with a variety of steady-state \cite{werdehausen2016coherent, mor_inhibition, nakano2018antiferroelectric, yan2019strong} and time-resolved \cite{werdehausen2016coherent, mor_inhibition, suzuki2020detecting, tang2020non, andrich2020imaging} methods. It was found that mode II is characterized by a very narrow lineshape in the low-temperature phase of Ta$_2$NiSe$_5$ \cite{werdehausen2016coherent} and undergoes a substantial broadening as the temperature approaches $T_C$ \cite{nakano2018antiferroelectric}. This is also confirmed by its apparent disappearance in time-domain data measured above $T_C$ \cite{werdehausen2016coherent,mor_inhibition} and its overdamped nature observed in spontaneous Raman scattering experiments \cite{kaiser}. Therefore, the long coherence time (and sharp linewidth) detected in our data of Fig. 3e is consistent with a scenario in which the crystal remains in the low-symmetry unit cell. This conclusion may be also supported by the presence of mode IV (at $\sim$3.7 THz) in our spectra. Previous studies found that below $T_C$ this mode lies in the vicinity of another phonon at 4.0 THz \cite{mor_inhibition, yan2019strong}. Both modes broaden and merge with increasing temperature and only one of them appears to persist above $T_C$. Therefore, it is not yet clear which mode (either the 3.7 THz or the 4 THz) is a characteristic phonon of the low-temperature phase of Ta$_2$NiSe$_5$. More refined resonant Raman scattering experiments with polarization selectivity and high energy resolution are needed to establish which mode disappears in the high-temperature orthorhombic phase. This aspect acquires particular importance in connection to the phonon calculations we present in Supplementary Note~7E. Our computation establishes that the soft phonon that drives the structural phase transition re-establishes around 3.5 and 4.0 THz in the low-temperature phase. In any event, for the sake of the current discussion, the sharp lineshape of mode II is a characteristic fingerprint of the low-temperature monoclinic structure of Ta$_2$NiSe$_5$, indicating the inhibition of the structural phase transition upon increase of $T_e$. This finding is confirmed by the direct visualization of the monoclinic distortion ($\beta$) angle as a function of time in UED (Fig. 4c) and expected from theoretical considerations (see Supplementary Note 7B).

\section*{Supplementary Note 7: First-principles calculations}

In this Supplementary Note, we discuss the results of our systematic \textit{ab initio} calculations.\\

\noindent \textbf{A. Relaxation of the unit cell}\\

Starting from the orthorhombic input geometry \cite{Jain2013, Ong_2015}, we
perform full structure relaxation of the lattice symmetry, cell volume,
and atomic coordinates. We include van der Waals forces using the
vdW-opt88-functional \cite{Klimes2010, Klime2011} and proceed with two
different structural relaxation methods: (i) by imposing the orthorhombic
symmetry, representative of the high-temperature phase, and (ii) by allowing
for a symmetry-free full relaxation, which results in a monoclinic cell
associated with the low-temperature phase. The relaxed angles for the latter
read $\alpha$~=~90.005$^\circ$, $\beta$~=~90.644$^\circ$, and $\gamma$ =
89.948$^\circ$. We note that the obtained monoclinic symmetry has a small
triclinic distortion, with an angle $\alpha$~=~90.005$^\circ$. These results
agree with the low-temperature structure determined experimentally, in which
the $\beta$ angle is between 90.53$^\circ$ and 90.69$^\circ$ \cite%
{sunshine1985structure,nakano2018antiferroelectric}. The slight deviation of the $%
\alpha$ angle from 90$^\circ$ would remain below the resolution of most
experimental methods for structural characterization. Finally, the lattice
constants for the orthorhombic cell after relaxation are $a$~=~3.512\AA , $b$~=~12.971\AA , $c$~=~15.778\AA , while for the monoclinic cell are $a$~=~3.517%
\AA , $b$~=~12.981\AA , $c$ = 15.777\AA.\\

\noindent \textbf{B. Effect of }$T_{e}$\textbf{\ on the structural distortion%
}\\

In this section, we provide an approximate estimate of the effect that $T_e$ has on
the structural distortion. We perform a series of calculations
in which we vary the Fermi smearing on the electronic eigenvalues. To quantify this
effect, we adapt the following algorithm. First, the structure is
optimized using the given Fermi smearing and the given density functional.
Then, we use the \textit{spglib} library \cite{togo2018texttt} to find a
structure with the orthorhombic symmetry $Cmcm$. Next, using the $maise$
library \cite{kolmogorov2012pressure}, we calculate the statistical distance between the original and the symmetrized structure, defined as%
\begin{equation}
d_{ab} = 1-c_{ab}/\sqrt{c_{aa}c_{bb}}.
\end{equation}%
In this expression, $c_{ab}$ reads
\begin{equation}
c_{ab} = \sum_{ij}\int R_{ij}^{a}(r)R_{ij}^{b}(r)dr,
\end{equation}
where $R_{ij}(r)$ is the standard radial distribution function (RDF) for
specific ionic species (in our case $i=\{$Ni, Ta, Se$\}$, which yields a total of 6 RDFs).

\begin{figure}
	\centering
	\includegraphics[width=0.85\columnwidth]{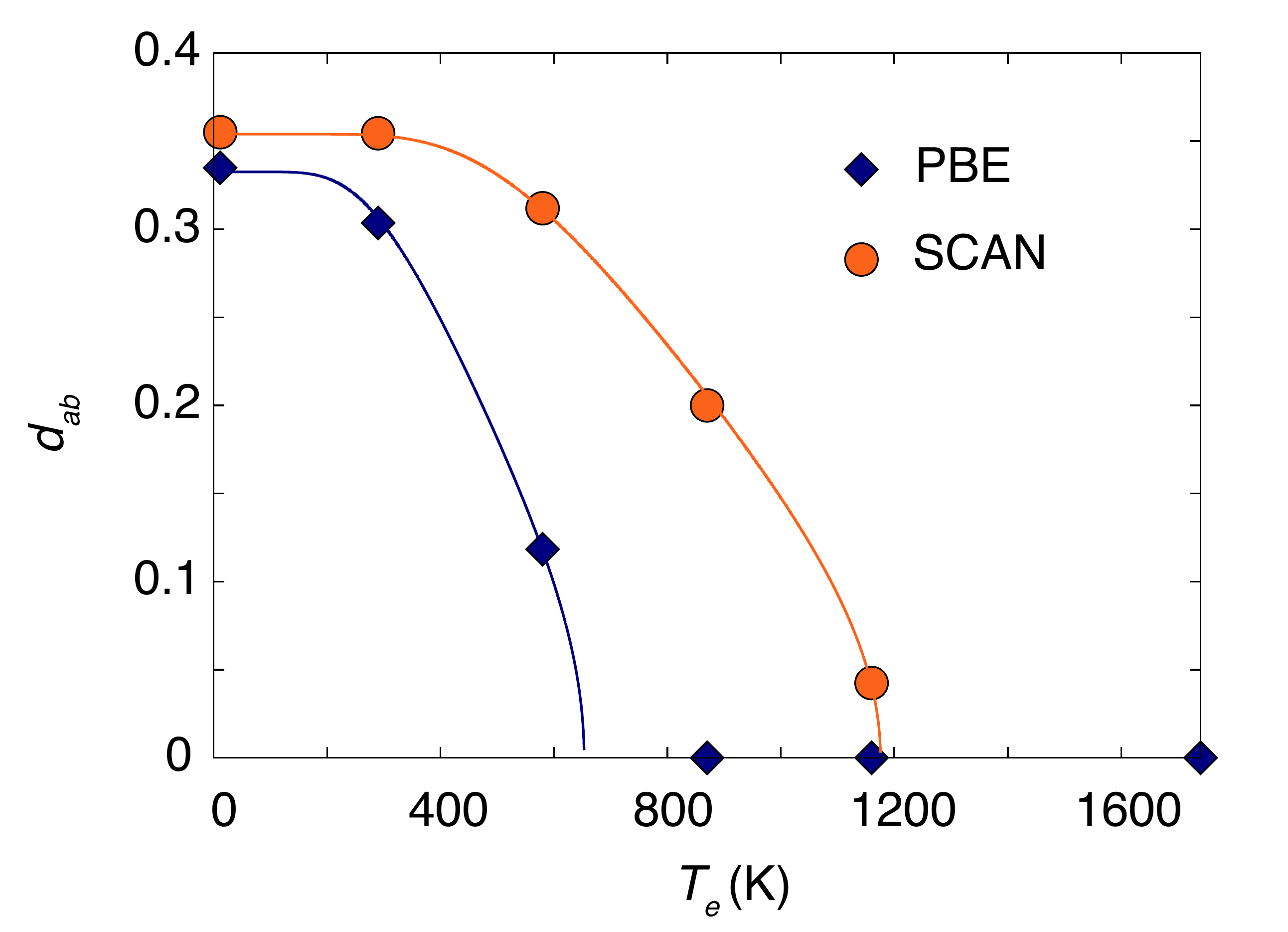}
	\caption{\textbf{Release of the monoclinic distortion upon increase of $T_e$.} Distance between the calculated monoclinic structure and the nearest orthorhombic structure, evaluated as a function of $T_e$. In the computation, we used two different density functionals: PBE (blue symbols) and SCAN (orange symbols). The lines are fits to the standard mean-field behavior, $d_{ab}(T_e)=d(0)\sqrt{\tanh(T_{c}/T_e)-1}$, where $T_C$ =~328~K.}
	\label{fig:distortion}
\end{figure}

In order to estimate the effect of the choice of the density functional, we
perform Density Functional Theory (DFT) calculations using the Perdew-Burke-Ernzerhof (PBE) functional \cite{perdew1996generalized} and the Strongly-Constrained and Appropriatly-Normed (SCAN) functional \cite{sun2015strongly}. The results are shown in Fig. \ref{fig:distortion}. While the zero-temperature structures hardly depend on the
choice of the functional, SCAN leads to about twice as high temperatures for the
structural instability. The main message of these results is that the structure remains monoclinic even at very high $T_e$, and the degree of monoclinic distortion depends of $T_e$. For the fluences used in the UED experiment, we see no change in the monoclinic $\beta$ angle within our resolution (Fig. 4c in the main text). This observation suggests that the SCAN functional may provide a more realistic description of the structural distortion release than the PBE functional. However, a quantitative comparison is hindered by the errors associated with the experimental estimate of the absorbed fluence and the evaluation of $T_e$ (see Supplementary Note 2).\\

\begin{figure}[t]
	\centering
	\includegraphics[width=0.75\columnwidth]{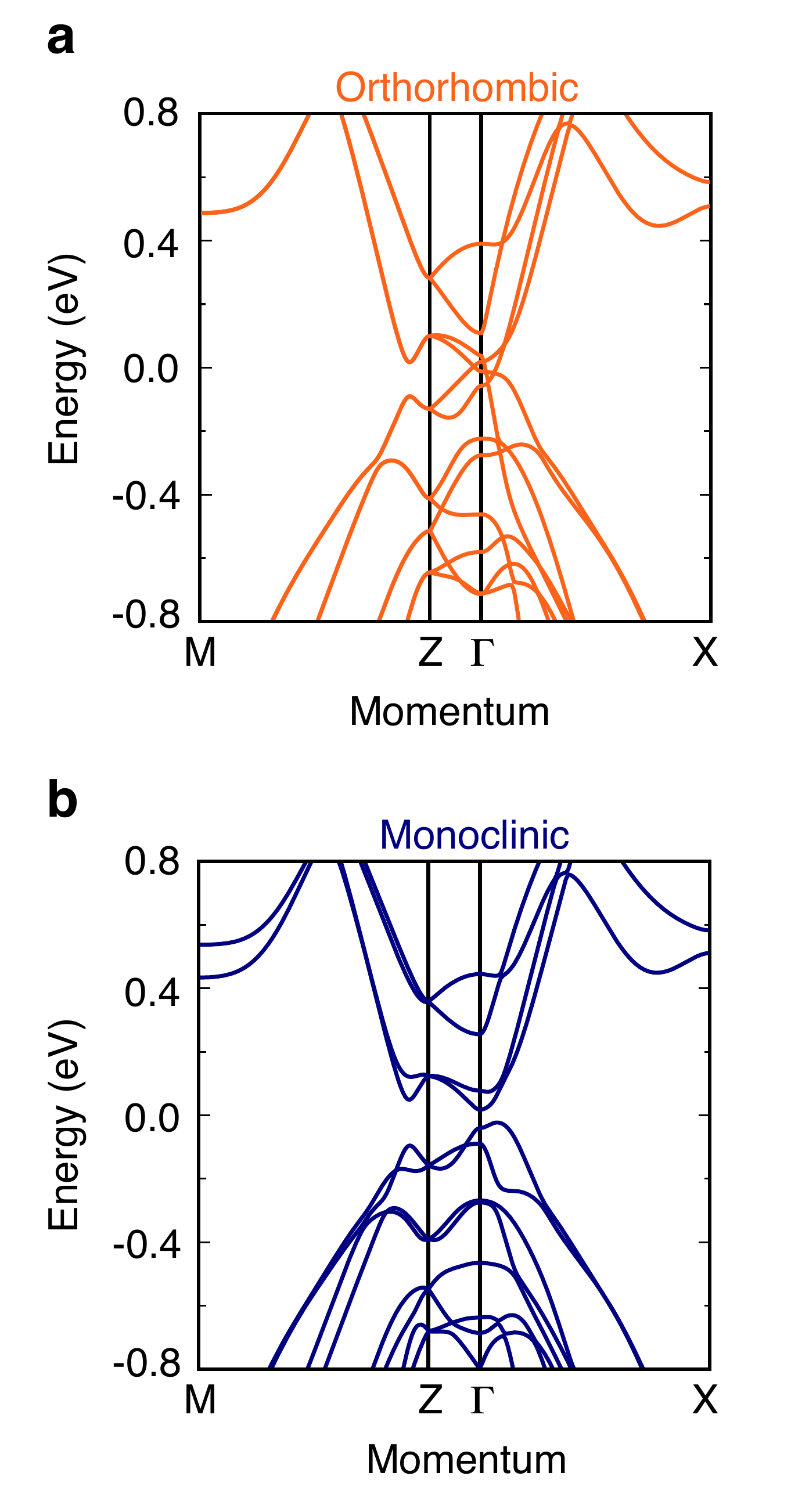}
	\caption{\textbf{DFT band structure calculations}. Electronic structure of Ta$_2$NiSe$_5$ calculated on the DFT level by using the PBE functional. \textbf{a}, The band structure in the orthorhombic lattice symmetry is semimetallic in nature. \textbf{b}, The band structure in the monoclinic lattice symmetry has a semiconducting character. This indicates that the structural transition alone opens a sizeable bandgap already on the level of standard DFT functionals. A similar effect was found using the modified Becke-Johnson functional.}
	\label{fig:DFT_bands}
\end{figure}

\noindent \textbf{C. Electronic structure at zero temperature}\\

To investigate the effect of the structural distortion on the electronic
structure of Ta$_{2}$NiSe$_{5}$, we perform DFT calculations for both the high-temperature orthorhombic and the
low-temperature monoclinic structures. After completing the structural
relaxation with the vdw-opt88-PBE functional \cite{Klimes2010, Klime2011}, we calculate the two
electronic structures with the PBE functional on a 16$\times$4$\times$4 $k$%
-mesh. The resulting band structures are shown in Fig.~\ref{fig:DFT_bands}.
We observe that the material is semimetallic in the orthorhombic phase, in agreement with previous results \cite{Kaneko2013}. In the orthorhombic structure, the VB and CB belong to two different irreducible representations of the crystal structure and no hybridization gap can be opened. This is why, in earlier studies on Ta$_2$NiSe$_5$ \cite{Kaneko2013}, the VB and CB of the orthorhombic phase were shifted artificially to account for the opening of a finite gap. Here instead we show, consistent with recent works \cite{watson2019band,subedi}, that performing the DFT calculations in the realistic monoclinic unit cell yields an indirect bandgap of 41 meV, since the VB and CB can lead to a finite hybridization gap in the lower-symmetry structure. Therefore, the structural phase transition alone causes bandgap opening and the material turns into a small-gap semiconductor. Moreover, already at the DFT level, the VB acquires an M-like dispersion around the $\Gamma $ point of the Brillouin zone,
similar to the experimental findings \cite%
{wakisaka2009excitonic,watson2019band}. In contrast, the CB dispersion does not accurately describe the measured one. Similar results are found by using different DFT functionals (e.g.,
the modified Becke-Johnson functional). We will discuss this aspect in a
separate long paper \cite{windgaetter}. In summary, our results demonstrate that the
change in crystal structure alone (i.e. without the need of including more
refined approximations for the electronic correlations) is sufficient to turn
Ta$_{2}$NiSe$_{5}$ from a semimetal into a small-gap semiconductor.

\begin{figure}
	\centering
	\includegraphics[width=1\columnwidth]{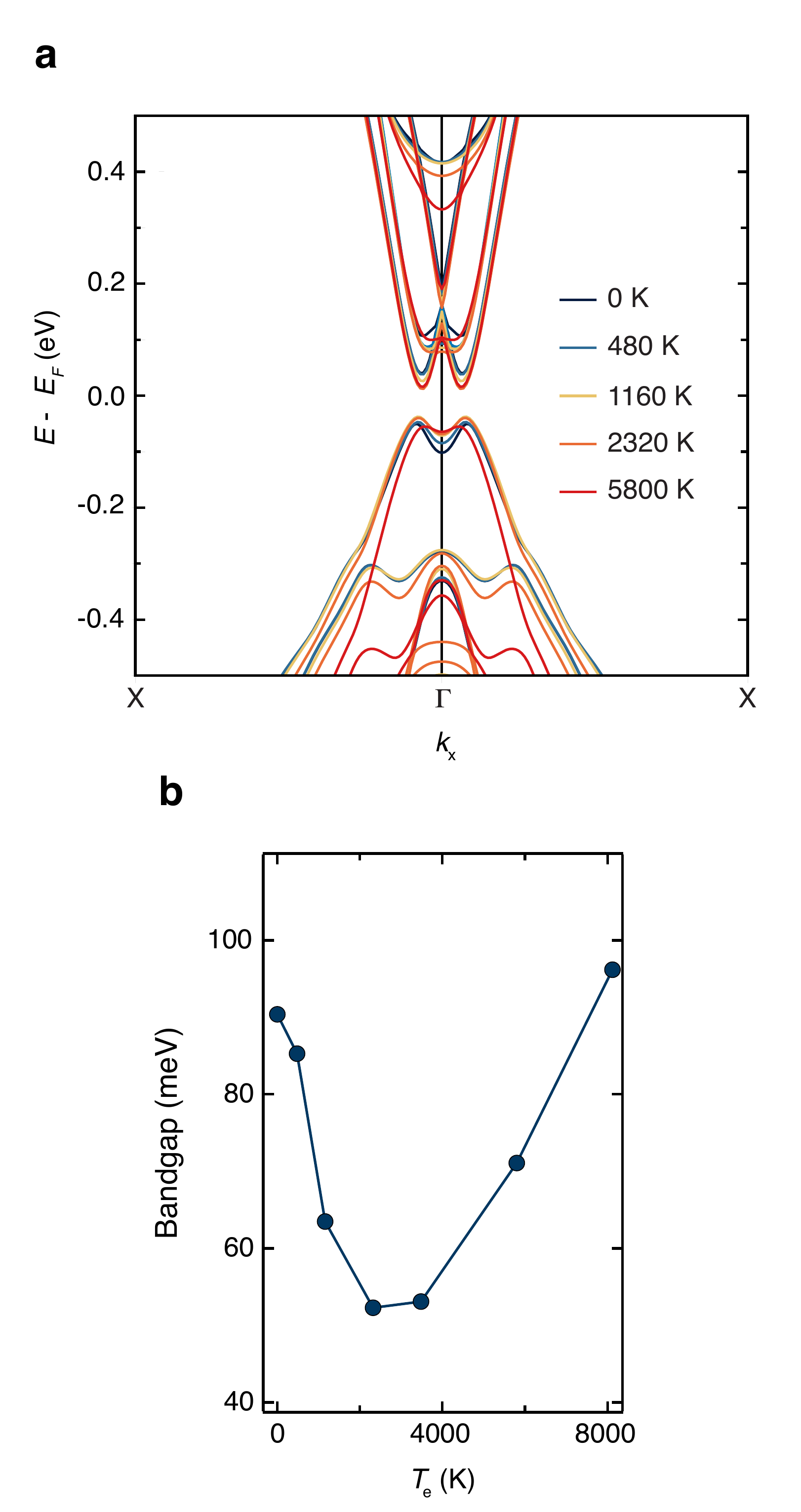}
	\caption{\textbf{Effect of $T_e$ on the electronic structure}.  \textbf{a},~Electronic structure of Ta$_2$NiSe$_5$ calculated on the $GW$ level as a function of $T_e$ (indicated with different color coding as specified in the label). \textbf{b}, Dependence of the size of the direct bandgap on the $T_e$. The gap shrinks compared to the equilibrium value for $T_e$$<$2500 K and starts to increase from the minimum value reached at $T_e$$>$2500 K. This behavior is consistent with the bandgap renormalization known in semiconductor physics \cite{spataru2004ab}. The calculations do not account, in principle, for the EI instability.}
	\label{fig:GW_bands_temperature}
\end{figure}

For a more accurate description of the electronic structure, we account
for many-body corrections within the $G_0W_0$ method \cite{hedin1965new}.
This level of theory neglects vertex-corrections for the self-energy and
terminates after one cycle of Hedin's equations. It has been proven to
refine the description of the band structures of semiconductors while being
still affordable computationally \cite{VanSchilfgaarde2006}. We perform the calculations in the monoclinic cell of Ta$_2$NiSe$_5$ using a
PBE-functional calculation as a starting point. The resulting band structure
is shown in Fig. 1b in the main text. We observe that the gap reaches a
value of 90 meV, with the VB maximum and CB minimum slightly shifted along
the $\Gamma$-X direction of the Brillouin zone. Spectroscopic ellipsometry experiments showed that the optical gap size at 10 K is $\sim$200 meV \cite{lu2017zero,larkin2018giant}, which is about twice the gap we find at the $GW$ level. STS data reported the equilibrium electronic gap to be $\sim$300 meV \cite{lee2019strong}. However, we note that in Ref.  \cite{lee2019strong} the value of the $\beta$ angle is not close to 90.5$^\circ$-90.6$^\circ$ value observed by diffraction techniques \cite%
{sunshine1985structure,nakano2018antiferroelectric}, but it is 92.5$^\circ$. To investigate the discrepancy between the reported equilibrium gaps, we iterate our $GW$ calculations for different monoclinic angles and find that the quasiparticle gap becomes larger with increasing $\beta$. Relying on this trend, we can extrapolate the value of the STS gap for the monoclinic angle expected for Ta$_2$NiSe$_5$. The final extrapolated gap is 165 meV. Thus, the 90 meV gap estimated in our $GW$ calculations for the ideal $\beta$ = 90.644$^\circ$ is a factor of 1.8 smaller than the experimental value. Further corrections in the description of the screening and the inclusion of the electron-phonon coupling in the $GW$ calculations would refine the gap size to larger values. The latter is already indicated by the frozen-phonon calculations we present in Supplementary Note 7F, where the most strongly-coupled mode driving the orthorhombic-to-monoclinic transition provides a 10 meV increase of the gap size.  The systematic
dependence of our $G_0W_0$ calculations on the initial conditions will be
also described in the separate paper \cite{windgaetter}. Besides the actual value of the gap, the most important feature obtained in our $GW$ calculations is the M-like (W-like) band dispersion that the VB (CB) acquires. This aspect is relevant because this characteristic dispersion has often been quoted in the literature as the fingerprint of the EI instability. In contrast to this scenario, the $GW$ method does not account for any ladder-diagrammatic effects such as the EI order and demonstrates that the structural degrees of freedom play a crucial role in determining the M-like (W-like) band flattening.\\

\noindent \textbf{D. Effect of $T_{e}$ on the electronic structure}\\

As anticipated in Supplementary Note 5, a pure EI (where the whole gap is opened by excitonic interactions) should respond to an increase of $T_e$ $\gg$ $T_C$ by showing complete closure of the gap. The origin of this effect lies in the screening of the Coulomb interaction caused by the photoexcitated carrier density (and therefore by $T_e$). This mechanism is also present under the name of bandgap renormalization in standard band semiconductors \cite{beni1978theory}, where the electronic structure is renormalized by the presence of a carrier population inside the bands. In this case, the bandgap typically shrinks at low carrier densities and increases at sufficiently high ones \cite{spataru2004ab}. In nonequilibrium experiments, bandgap renormalization manifests itself promptly with photoexcitation because it is an electronic process in nature, governed by the plasma frequency. In our experimental data, we observe the gap size to respond promptly to photoexcitation (Supplementary Note 5C and Fig. \ref{fig:Lorentz}e), but this effect is small compared to that caused by the structural degrees of freedom. As such, the gap of Ta$_2$NiSe$_5$ size remains large at all time delays, in contrast to the scenario expected for a pure EI. In this section, we investigate theoretically how the electronic structure of the monoclinic phase of Ta$_2$NiSe$_5$ changes upon an increase of $T_e$ and quantify the strength of carrier-induced screening effects.

\begin{figure}[b]
	\centering
	\includegraphics[width=0.85\columnwidth]{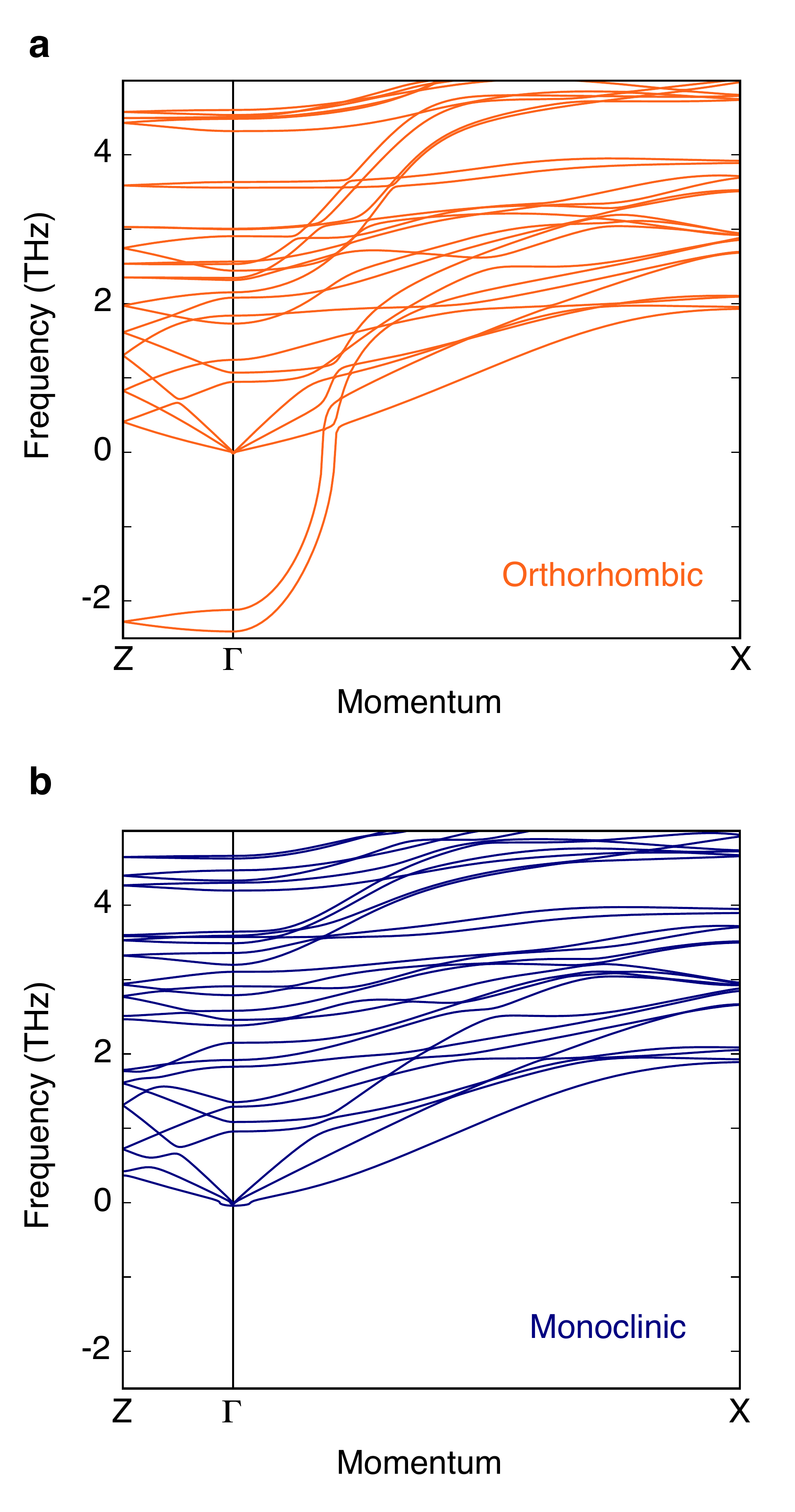}
	\caption{\textbf{Phonons in Ta$_2$NiSe$_5$.} Phonon energy-momentum dispersion curves calculated in the \textbf{a}, orthorhombic and \textbf{b}, monoclinic cell. Two modes acquire a negative (i.e. imaginary) frequency in the orthorhombic structure around the $\Gamma$ point of the Brillouin zone, which indicates that these phonons are soft modes of the high-temperature unit cell of Ta$_2$NiSe$_5$. The modes reappear in the phonon dispersion of the low-temperature monoclinic cell around a frequency of 3.5-4 THz.}
	\label{fig:phonon_bands}
\end{figure}

We perform $GW$ calculations by assuming a thermalized distribution of electrons
(holes) in the CB (VB), set through the Fermi-Dirac distributions for
different values of $T_e$ around a value of $E_F$ that is determined self-consistently. Specifically, we set a Fermi smearing corresponding to different values of $T_e$ (spanning the range 0-8000 K) at the beginning of the self-consistent DFT calculations. Afterwards,
we perform $G_0W_0$ calculations on top of such DFT ground states. The
results are shown in Fig. \ref{fig:GW_bands_temperature}a for representative value of $T_e$. 

\begin{figure*}[t]
	\centering
	\includegraphics[width=1.5\columnwidth]{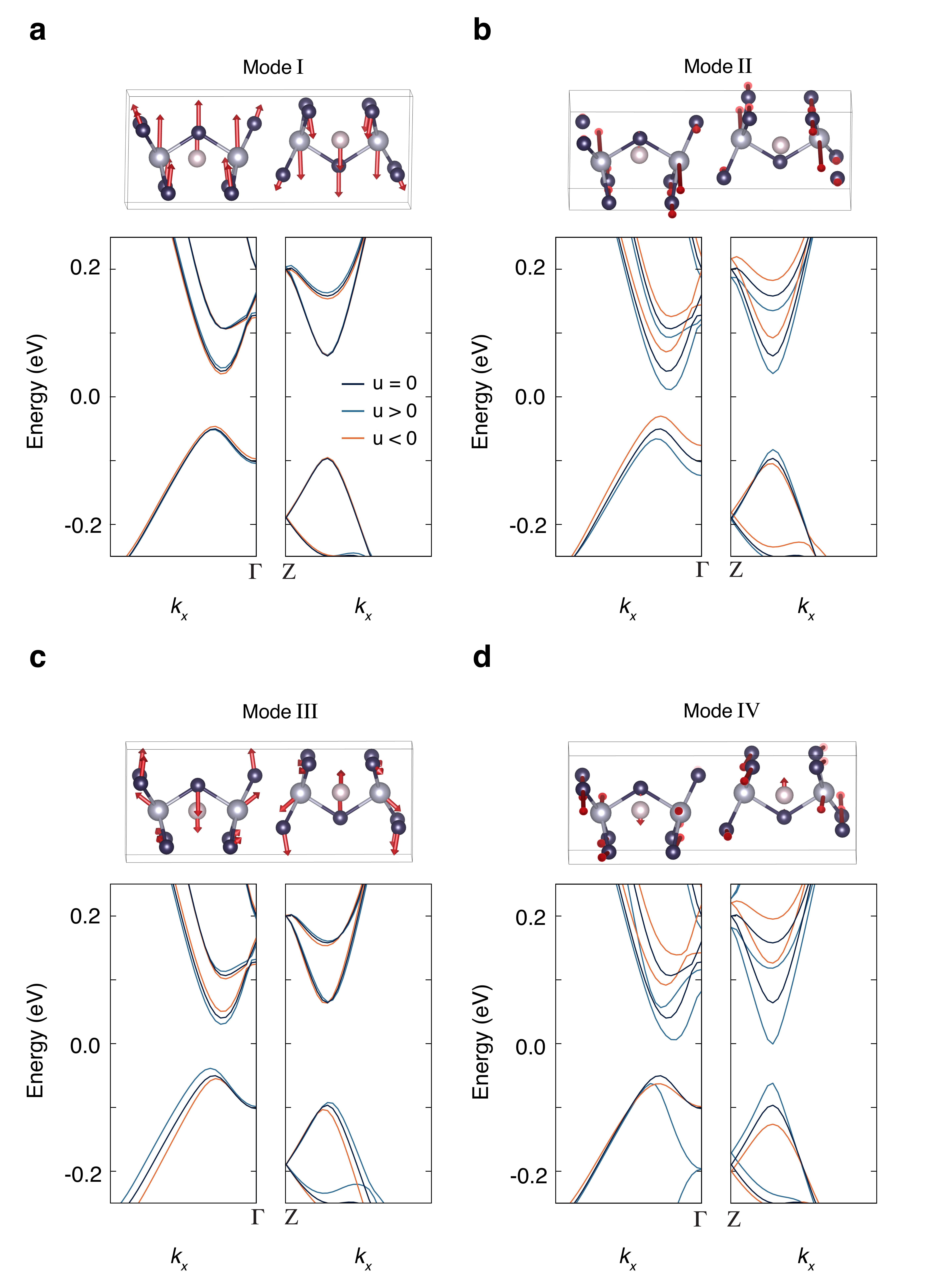}
	\caption{\textbf{Effect of the phonon displacements on the electronic structure}. \textbf{a-d}, Calculated eigenvectors of the dynamical matrix of Ta$_2$NiSe$_5$ corresponding to modes (a) I, (b) II, (c) III, (d) IV of Fig. 3e. Light-blue atoms refer to Ta, pink atoms to Ni, and dark blue atoms to Se (see the legend in Fig. 3 in the main text). The phonon spectrum has been computed using DFT. To enhance the visibility of the atomic motion, the amplitude is scaled by a factor of 8. The electronic structure of Ta$_2$NiSe$_5$ displaced along the eigenvectors of each modes is also shown. The dashed grey lines refer to the electronic structure of the initial (undisplaced) low-temperature unit cell, whereas the orange (blue) solid lines indicate the band structure for positive (negative) displacements. The electronic structures are computed on the $GW$ level of theory.}
	\thispagestyle{empty}
	\label{fig:phonon_displaced}
\end{figure*}

We observe that at $T_e$ comparable to those produced by photoexcitation in our experiments ($<$ 1160 K, see Supplementary Note 2), the band
structure of Ta$_2$NiSe$_5$ displays a very small modification. The most pronounced change occurs at the top of the VB, where the M-like shape is slightly released and a more flattened dispersion develops. At high momenta, no variation in the dispersion occurs. Only when $T_e$~$>$~4000~K (i.e. for unrealistic $T_e$ in experiments, as they lie above the damage threshold of the material), the band structure renormalizes in a significant way. One important effect is that the temperature-corrections are quite anisotropic and larger for the CB than the VBs. We also study how the direct bandgap of Ta$_2$NiSe$_5$ reacts to the increase in $T_e$ (Fig. \ref{fig:GW_bands_temperature}b). We see that at low values of $T_e$ (0~$<$~$T_e$~$<$~1160~K), the gap shrinks by a small amount (5-27 meV). At $T_e$ $\sim$ 2500 K, the gap reaches its minimum value and at higher $T_e$ an opposite trend starts, consistent with the well-known temperature-dependent bandgap renormalization effect of standard semiconductors \cite{spataru2004ab}. Finally, we remark that in our calculations the chemical potential for a given temperature is fixed by the Fermi Dirac distribution, while in the experiments the slight doping of the samples makes the chemical potential pinned closer to the CB \cite{lee2019strong}. By considering all these aspects, we can rationalize the prompt response of VB top observed in our trARPES experiments.\\

\noindent \textbf{E. Phonon energy-momentum dispersion}\\

Since our experimental results point toward a substantial contribution of the lattice degrees of freedom in the phase transition of Ta$_2$NiSe$_5$, here we investigate the lattice dynamics in the material's orthorhombic and monoclinic unit
cells using Density Functional Perturbation Theory \cite%
{DFPT} with the vdw-opt88-PBE functional \cite{Klimes2010, Klime2011}. The calculated phonon dispersions are
displayed in Fig. \ref{fig:phonon_bands}. In the orthorhombic structure
(Fig.~\ref{fig:phonon_bands}a) we observe two phonon modes with imaginary
frequency, in agreement with the results of Ref. \cite{subedi}. The presence
of imaginary frequencies is a signature that these phonons are soft modes of
the orthorhombic unit cell and drive the structural phase transformation. As
expected, these two modes stabilize after full relaxation in the
low-temperature structure (Fig. \ref{fig:phonon_bands}b), acquiring a
frequency of $\sim$3.5~THz. We analyze the eigenmodes of the phonons
emerging in our trARPES experiment using the $Phonopy$ package \cite{phonopy}.
The phonon eigenvectors are displayed in Fig.~\ref{fig:phonon_displaced} and
show agreement with those analyzed previously in the literature \cite%
{werdehausen2016coherent}. In particular, mode IV is the mode that
originates from the soft phonon driving the orthorhombic-to-monoclinic
transition. As such, its eigenvector involves the shear motion of the
neighboring Ta chains around the Ni chain.\\

\noindent \textbf{F. Electronic structure in a frozen-phonon unit cell}\\

We also analyze the effect that phonon modes I-IV have on the electronic
band structure of Ta$_2$NiSe$_5$. We perform $G_0W_0$ calculations while
statically displacing the atoms in the unit cell along specific phonon
eigenmodes. While this adiabatic method can provide information only on
the electron-phonon coupling in the electronic ground state \cite{baldini2020electron}, it represents
a first important step to elucidate how specific atomic motions affect the
electronic properties of Ta$_2$NiSe$_5$. Our results are shown in Fig. \ref%
{fig:phonon_displaced}, where we present the band structure along the $\Gamma
$-X and Z-M momentum directions for modes I-IV. Note that, to enhance the
visibility of the phonon-induced changes, the energy axis is not
aligned around $E_F$, but it is referenced to the energy of infinitely
separated atoms. We observe that the low-energy VB states around the $\Gamma
$ point are more sensitive toward the displacement of mode II, in agreement
with our results of Fig. 3e. Mode IV leads to an asymmetric and nonlinear behavior of the band structure upon the application of positive and the negative displacements. This behavior occurs in the presence of a non-perturbative strong electron-phonon coupling, which breaks the linear and symmetric dependence expected from the traditional deformation potential theory. Finally, repeating similar calculations at different
temperatures (100~K and 300~K) yields no substantial changes in our results.

%\bibliography{Papers}

\providecommand{\noopsort}[1]{}\providecommand{\singleletter}[1]{#1}%

\end{document}